\documentclass[
  prc,notitlepage,twocolumn,showpacs,floatfix,nofootinbib,
  preprintnumbers,superscriptaddress,aps,longbibliography,10pt
]{revtex4-2}

\usepackage{graphicx}
\usepackage{amsmath,amssymb}
\usepackage{bm}
\usepackage{color}
\usepackage{dcolumn}
\usepackage{enumerate}
\usepackage{multirow}
\usepackage{threeparttable}
\usepackage{mathrsfs}
\graphicspath{{figs/}}

\usepackage{silence}
\usepackage[bookmarks=true,pdfpagelabels, pdfencoding=auto, psdextra]{hyperref}
\hypersetup{
  pdfsubject=Paper,
  pdfkeywords={Nuclear Physics} {Density Functional Theory} {Mean Field Theory} {QRPA}  {reduced basis method} {emulators} ,
  unicode = true,
  breaklinks = true,
  colorlinks = true,
  linkcolor = blue,
  citecolor = blue,
  menucolor = blue,
  urlcolor = blue
}
\usepackage{orcidlink}
\begin{document}

\title{Reduced-basis method for linear response within
nuclear density functional theory}
\author{Nobuo Hinohara~\orcidlink{0000-0001-9562-0189}}
\email{hinohara@nucl.ph.tsukuba.ac.jp}
\affiliation{
 Center for Computational Sciences, \href{https://ror.org/02956yf07}{University of Tsukuba}, Tsukuba, 305-8577, Japan
}
\affiliation{
 Faculty of Pure and Applied Sciences, \href{https://ror.org/02956yf07}{University of Tsukuba}, Tsukuba, 305-8571, Japan
}
\affiliation{\href{https://ror.org/03r4g9w46}{Facility for Rare Isotope Beams}, \href{https://ror.org/05hs6h993}{Michigan State University}, East Lansing, MI~48824, USA}

\author{Xilin Zhang~\orcidlink{0000-0001-9278-5359}} 
 \email{zhangx@frib.msu.edu}
\affiliation{\href{https://ror.org/03r4g9w46}{Facility for Rare Isotope Beams}, \href{https://ror.org/05hs6h993}{Michigan State University}, East Lansing, MI~48824, USA}

\author{Jonathan Engel~\orcidlink{0000-0002-2748-6640}}
\email{engelj@physics.unc.edu}
\affiliation{
Department of Physics and Astronomy, \href{https://ror.org/0130frc33}{University of North Carolina}, Chapel Hill, North Carolina, 27599-3255, USA
}

\date{\today}

\begin{abstract}
\begin{description}
\item[Background]
The quasiparticle random-phase approximation (QRPA) within nuclear density functional theory provides a powerful framework for describing collective excitations. 
Although the finite-amplitude method (FAM) provides efficient access to QRPA solutions,
it is still 
computationally demanding when the FAM equations have to be solved multiple times by varying the parameters of the external field. 
Reduced-order emulators, such as the reduced-basis method (RBM), offer a promising strategy for accelerating repeated linear-response calculations.
\item[Purpose]
We aim to construct an RBM-based emulator for the FAM that treats the complex frequency (energy) of the external field as a model parameter. 
The goal is to efficiently reproduce the FAM amplitudes and approximate QRPA eigenmodes in the energy region of interest.
\item[Methods]
High-fidelity FAM calculations are performed at a small set of training energies in the complex-energy plane. 
The FAM amplitudes at training energies are used as non-orthogonal basis functions to construct a reduced-order representation of the FAM equation. 
The variational equation yields an emulator that can predict not only the response at arbitrary complex energies,  but also the QRPA eigensolutions, without additional full FAM calculations. 
\item[Results]
The RBM emulator accurately reproduces the FAM strength distribution both in the giant-resonance region and for low-energy QRPA states when the relevant energy domain is covered by the training set. 
The emulator also reproduces the imaginary QRPA modes associated with shape instabilities of the HFB state. 
The method is applied to the $K^\pi=0^+$ mode of rare-earth Dy isotopes in a realistic model space, and the RBM emulator describes both the strength distributions and the lowest $0^+$ collective states with precision comparable to that of full FAM calculations, reducing the computational cost by more than an order of magnitude.
\item[Conclusions]
The RBM provides an efficient and accurate emulator for FAM calculations. By drastically reducing the computational cost while preserving high-fidelity reproduction of both giant-resonance and low-energy modes, including imaginary-energy solutions, the RBM emulator can be used to optimize 
density-functional parameters, calculate collective inertia, and conduct large-scale surveys of nuclear collective excitations.
\end{description}
\end{abstract}

\maketitle

\section{Introduction \label{sec:intro}}

Nuclear density functional theory (DFT) provides a microscopic framework for describing nuclear ground states \cite{10.1088/2053-2563/aae0ed}.
There exist many global calculations of ground-state properties across the nuclear chart, with a variety of energy density functionals (EDFs) \cite{PhysRevC.68.054312,ZHANG2022101488,PhysRevC.108.044316}.
The time-dependent extension of nuclear DFT (TDDFT) allows the description of excited states and nuclear dynamics, including nuclear collisions, but is far more computationally intensive than static DFT \cite{RevModPhys.88.045004}.

The quasiparticle random-phase approximation (QRPA) \cite{Ring-Schuck,Blaizot-Ripka} is a small-amplitude approximation to nuclear TDDFT that is applied widely to obtain, e.g., strength distributions around giant resonances, energies of and strengths to low-lying excited states, $\beta$-decay rates, and rotational and pairing-rotational moments of inertia.
The QRPA is also a building block for more complex theories such as the QRPA-plus-quasiparticle-vibration coupling \cite{PhysRevC.109.044308}, the second QRPA, time-dependent density matrix theory \cite{10.3389/fphy.2020.00067}, the local QRPA \cite{PhysRevC.82.064313}, and the adiabatic self-consistent collective coordinate method \cite{PTP.103.959}.
Despite their relative simplicity, the QRPA equations with a realistic DFT are still computationally demanding, especially when the nuclear ground state is deformed and rotational symmetry is broken.
The QRPA phonons that connect the ground and excited states are a superposition of two-quasiparticle creation and annihilation operators.
Conventionally, the QRPA equations are solved by matrix diagonalization in the space of such states (in the quasi-boson approximation) together with a truncation of the two-quasiparticle model space.
The iterative Lanczos \cite{Johnson1999155} and Arnoldi 
methods \cite{PhysRevC.81.034312,PhysRevC.86.014307,PhysRevC.86.024303} have also been developed to effectively reduce the size of the two-quasiparticle basis space.

The finite-amplitude method (FAM) for QRPA \cite{nakatsukasa:024318,PhysRevC.84.014314,EPJA.62.133} was proposed as an iterative solution of the TDDFT linear-response equations.
The FAM solves these equations, which contain a one-body time-dependent external field, as forced oscillations.
Because the FAM does not require the construction and diagonalization of the QRPA matrix, it handles the large two-quasiparticle space without additional truncation at the QRPA level.  
Based on this formulation, iterative schemes have been developed to calculate the FAM response function over a broad energy range, by expanding it in Chebyshev polynomials in conjunction with the kernel polynomial method \cite{BJELCIC2022108477} or by reducing it to a Krylov subspace with a Lanczos method \cite{roh2026lanczosmethodqrpastrength}.
The FAM has been successfully applied to giant resonances \cite{PhysRevC.93.034329}, and together with contour-integration techniques \cite{PhysRevC.87.064309,PhysRevC.91.044323}, to the computation of $\beta$-decay rates in thousands of isotopes \cite{PhysRevC.93.014304,PhysRevC.94.055802,PhysRevC.102.034326} and of the matrix elements that govern two-neutrino double-beta decay \cite{PhysRevC.105.044314}.

One would like to use the method in ways that require even faster computation, however.  To use, e.g, giant-resonance energy centroids to optimize density functionals, to compute the collective inertia on multi-dimensional potential-energy surfaces associated with shape coexistence or fission dynamics \cite{PhysRevC.109.L051301,PhysRevC.103.014306}, or to calculate the matrix elements for neutrinoless double-beta decay 
with non-separable two-body operators \cite{PhysRevC.87.064302}, one must obtain many 
low-energy QRPA solutions with good precision.  
Doing that requires numerous FAM calculations in which one varies the parameters of the problem, e.g.\ the coupling constants in density functionals, the constrained deformation parameters, or the parameters that describe the non-separable decay operators.

A reduced-order emulator would help.  Such emulators \cite{Melendez_2022,10.3389/fphy.2022.1092931,RevModPhys.96.031002} can provide approximate solutions to high-fidelity models very quickly.
The reduced-basis method (RBM) is suitable for parameter-dependent problems and for situations in which equations must be solved repeatedly while varying parameters.
The basic idea of the RBM is to use high-fidelity solutions for a few values of the parameters to construct an approximate reduced-order space of solutions. 
Then that space is used to rapidly obtain the solutions for different values of the parameters.
Eigenvector continuation, a version of the method, 
has been applied extensively in nuclear physics, e.g.\ to
quantum Monte Carlo simulations \cite{PhysRevLett.121.032501}
and shell model calculations \cite{10.1093/ptep/ptac057}.
Other forms of the RBM have been used in nuclear DFT \cite{PhysRevC.106.054322} and for scattering problems \cite{FURNSTAHL2020135719,PhysRevC.105.064004}. A different class of emulators, exemplified by the parametric matrix model (PMM)~\cite{Cook:2024toj}, retains the algebraic structure of an eigenvalue problem but treats the elements of the reduced matrices as free parameters to be fitted to high-fidelity data. The RBM used here is instead model driven: the reduced matrices are obtained by projecting the FAM equation onto the subspace of high-fidelity solutions, without adjustable parameters.

In electronic structure theory, an RBM has been applied to linear response \cite{doi:10.1021/acs.jctc.7b00402}. 
In this article, we construct an RBM emulator for the FAM-QRPA, which produces the nuclear linear response. 
As do emulators for continuum physics \cite{PhysRevLett.135.242501,PhysRevC.112.064605}, we regard the complex frequency (energy) of the applied external field as a model parameter
and emulate both the strength function over the complex-energy plane and the locations and residues of the QRPA poles.
We demonstrate the power of the emulator, both for 
strength functions and low-energy QRPA states.  We also compare our emulator to
the iterative Arnoldi method, showing that the emulator 
has the advantage of being able to enhance resolution in any particular energy region.

The paper is organized as follows:
In Sec.~\ref{sec:formalism}, after briefly recapitulating the QRPA and the FAM, we develop an RBM for the FAM.
In Sec.~\ref{sec:results}, we analyze a schematic test in detail.
Section~\ref{sec:beta} presents realistic examples: systematic calculations of isoscalar giant quadrupole resonances and $\beta$ vibrations in the rare-earth region.
Section~\ref{sec:conclusion} is the conclusion.

\section{formalism \label{sec:formalism}}

\subsection{QRPA}

The QRPA describes both the ground and excited states in terms of quantized mean-field oscillations~\cite{Ring-Schuck, Blaizot-Ripka}.
The computational challenge is solving a non-Hermitian eigenvalue problem of the following form
\begin{align}
\begin{pmatrix} A & B \\ -B^\ast & -A^\ast \end{pmatrix}
\begin{pmatrix} X^\lambda \\ Y^\lambda \end{pmatrix}
= 
\Omega_\lambda
\begin{pmatrix} X^\lambda \\ Y^\lambda \end{pmatrix}. \label{eq:QRPAeq}
\end{align}
The form of the QRPA equation allows complex eigenvalues \cite{doi:10.1093/ptep/ptw073}.
Throughout this section, however, to avoid unnecessary complexity, we consider only cases in which the eigenvalue $\Omega_\lambda$ is a nonzero real number. 
We summarize the situation with imaginary $\Omega_\lambda$ separately in Appendix~\ref{sec:PQrepresentation}. 
The positive eigenvalue $\Omega_\lambda>0$ represents the energy of the $\lambda$-th QRPA excited state, and the components $X^\lambda$ and $Y^\lambda$ of the corresponding eigenvector are the two-quasiparticle amplitudes that specify the structure of that state. A phonon operator constructed from annihilation operators $\hat{a}$ and creation operators $\hat{a}^\dag$ in the quasiparticle basis\footnote{The quasiparticles define the Hartree-Fock-Bogoliubov (HFB) mean-field solution as a vacuum through the relation 
$\hat{a}|{\rm HFB}\rangle=0$.},
\begin{align}
{\cal O}^\dag_\lambda = \sum_{\mu<\nu} X_{\mu\nu}^\lambda\hat{a}^\dag_\mu\hat{a}^\dag_\nu
- Y_{\mu\nu}^\lambda\hat{a}_\nu \hat{a}_\mu,
\end{align}
creates the excited state $|\lambda\rangle$ from the QRPA ground state $|0\rangle$:
\begin{align}
    |\lambda\rangle = {\cal O}^\dag_\lambda |0\rangle  \,.
\end{align}

The matrices $A$ and $B$ in Eq.\ (\ref{eq:QRPAeq}) are Hermitian and symmetric, respectively. 
They contain the residual interaction and are given by second functional derivatives of the EDF with respect to the quasiparticle density matrix; see, e.g., Ref.~\cite{Blaizot-Ripka}.
The basis in which the matrices are written consists of two-quasiparticle states with quasiparticle labels $\mu$ and $\nu$.
The dimension of the single-quasiparticle space is 1,771 in oscillator shells
with $N_{\rm max} = 20$,
if time-reversal and axial symmetries are conserved, and the number of two-quasiparticle basis states (the dimension of the $A$ and $B$ matrices) is 128,843 for $K^\pi=0^+$ for axial- and reflection-symmetric HFB ground states.
In diagonalizing the QRPA matrix, one sometimes imposes a cutoff on the two-quasiparticle energy \cite{PhysRevC.83.021304,Yoshida01112013} or the single-particle occupation probability \cite{PhysRevC.82.034326,PhysRevC.84.014332,PhysRevC.87.064302} to reduce the number of two-quasiparticle states. 

If the eigenvalue and eigenvector in the set ($\Omega_\lambda, X^\lambda, Y^\lambda)$ 
are a solution of the QRPA equation, then those in
$(-\Omega_\lambda, Y^{\lambda\ast}, X^{\lambda\ast})$ are also a solution \cite{doi:10.1093/ptep/ptw073}.
Using that fact, one can write the QRPA equations in the form
\begin{align}
{\cal S}{\cal X} = \Sigma_3 {\cal X}{\cal O},
\end{align}
where the matrices are defined by
\begin{subequations}
\begin{align}
{\cal S} &= \begin{pmatrix} A & B \\ B^\ast & A^\ast \end{pmatrix}, &\quad
{\cal X} &= \begin{pmatrix} X & Y^\ast \\ Y & X^\ast \end{pmatrix} \label{eq:stabilitymat} \\
\Sigma_3 &= \begin{pmatrix} 1 & 0 \\ 0 & -1 \end{pmatrix}, &\quad
{\cal O} &= \begin{pmatrix} \Omega & 0 \\ 0 & -\Omega \end{pmatrix} \,,
\end{align}
\end{subequations}
with $\Omega_{ij} = \Omega_i \delta_{ij}$.
The QRPA eigenvectors are normalized through the conditions, 
\begin{align}
    {\cal X}^\dag \Sigma_3{\cal X} = \Sigma_3, \quad
    {\cal X}\Sigma_3{\cal X}^\dag = \Sigma_3 \,. \label{eq:XYnormalization}
\end{align}

\subsection{FAM}

Diagonalizing the QRPA matrix in the full two-quasiparticle space is numerically demanding. We are usually 
interested in a small number of quantities obtained from the eigenvalues and eigenvectors of the QRPA, e.g.\ the 
wave functions and/or excitation energies of a few low-lying states or a Lorentzian-smeared strength distribution as a function of the excitation energy.
The FAM is a version of the QRPA that is based on linear-response theory
(the small-amplitude limit of TDDFT)~\cite{nakatsukasa:024318,PhysRevC.84.014314}.
In the FAM, a time-dependent external field of the form
\begin{align}
 \hat{F}(t) = \hat{F}e^{-i\omega t} + \hat{F}^\dag e^{i\omega t} \,,
\end{align}
is applied to the system.  Here
$\hat{F}$ is a one-body operator and $\omega$ is a frequency.
TDDFT produces the linear-response equation
\begin{align}
\left[ 
{\cal S}
- \omega
\Sigma_3 \right]
\begin{pmatrix} X(\omega) \\ Y(\omega) \end{pmatrix}
=
- \begin{pmatrix} F^{20} \\ F^{02} \end{pmatrix} \,,  \label{eq:FAMeq}
\end{align}
where $F^{20}$ and $F^{02}$ are the two-quasiparticle matrix elements of the operator $\hat{F}$:
\begin{subequations}
\begin{align}
F^{20}_{\mu\nu} &= \langle {\rm HFB} | \hat{a}_\nu \hat{a}_\mu \hat{F} | {\rm HFB}\rangle \,, \\
F^{02}_{\mu\nu} &= \langle {\rm HFB} | \hat{F} \hat{a}_\mu^\dag \hat{a}_\nu^\dag | {\rm HFB}\rangle \,.
\end{align}
\end{subequations}
The frequency $\omega$ (hereafter referred to as the complex energy) is regarded as a complex parameter in Eq.~(\ref{eq:FAMeq}).
The solutions of that equation are the amplitudes $X(\omega)$ and $Y(\omega)$ (representing the system's response) as a function of the external field $\hat{F}$ and $\omega$. The amplitudes are related to the $X^\lambda$ and $Y^\lambda$ of Eq.~\eqref{eq:QRPAeq} via Eq.~\eqref{eq:FAMXY} below. 
An important advantage of Eq.~(\ref{eq:FAMeq}) is that it can be solved without explicitly computing and storing the $A$ and $B$ matrices. First, one notes that the matrix-vector operations $AX(\omega)+BY(\omega)$ and $B^\ast X(\omega)+A^\ast Y(\omega)$ in Eq.~\eqref{eq:FAMeq} lead to a change in Hamiltonian fields:
\begin{subequations}
\begin{align}
\delta H^{20}_{\mu\nu}(\omega) &= 
[AX(\omega)+BY(\omega)]_{\mu\nu} -(E_\mu+ E_\nu)X_{\mu\nu}(\omega) , \\
\delta H^{02}_{\mu\nu}(\omega) &= 
[A^\ast Y(\omega) + B^\ast X(\omega)]_{\mu\nu} -(E_\mu+E_\nu)Y_{\mu\nu}(\omega) \,.
\end{align}
\end{subequations}
Here $E$ represents the quasiparticle energy.  
Equations (\ref{eq:FAMeq}) then take the simple FAM-equation form
\begin{subequations}
\label{eq:SimpFAM}
\begin{align}
 X_{\mu\nu}(\omega) &= - \frac{ \delta H^{20}_{\mu\nu}(\omega) + F^{20}_{\mu\nu}}{ E_{\mu} + E_{\nu} - \omega},\\
 Y_{\mu\nu}(\omega) &= - \frac{ \delta H^{02}_{\mu\nu}(\omega) + F^{02}_{\mu\nu}}{ E_{\mu} + E_{\nu} + \omega} \,.
 \end{align}
\end{subequations}
As shown, e.g.\ in Ref.~\cite{PhysRevC.84.014314}, the induced fields $\delta H^{20}(\omega)$ and $\delta H^{02}(\omega)$ can be computed directly from the nuclear EDF, allowing one to avoid $A$ and $B$ altogether.
Because the induced fields are functions of $X(\omega) $ and $Y(\omega)$, one typically solves the above equations iteratively.

The FAM equations provide a formal relation between the FAM amplitudes $X(\omega), Y(\omega)$ and the QRPA eigenvectors
\cite{Blaizot-Ripka,PhysRevC.87.064309,PhysRevC.92.034321}.
We present the decomposition of the matrix in Eq.~(\ref{eq:FAMeq}) for the case when only real and nonzero eigenvalues are present (see Appendix~\ref{sec:PQrepresentation} for the more general situation):
\begin{align}
[ {\cal S} - \omega \Sigma_3]^{-1} &=
{\cal X}[{\cal O} - \omega]^{-1} \Sigma_3 {\cal X}^\dag \,.
\end{align}
The FAM amplitudes have the following spectral decomposition in terms of the QRPA eigenmodes: 

\begin{subequations}
\begin{align}
    X_{\mu\nu}(\omega) 
    &= - \sum_{\lambda} \biggr[
    \frac{ X^\lambda_{\mu\nu}\langle \lambda |\hat{F}|0\rangle }{\Omega_\lambda - \omega} +
    \frac{ Y^{\lambda\ast}_{\mu\nu}\langle 0 |\hat{F}|\lambda\rangle }{\Omega_\lambda + \omega}\biggr], \\
Y_{\mu\nu}(\omega) &= -\sum_{\lambda}
    \biggr[\frac{ Y^\lambda_{\mu\nu}\langle \lambda |\hat{F}|0\rangle }{\Omega_\lambda - \omega} +
    \frac{ X^{\lambda\ast}_{\mu\nu}\langle 0 |\hat{F}|\lambda\rangle }{\Omega_\lambda + \omega}\biggr] \,,
\label{eq:FAMXY}
\end{align}
\end{subequations}
where the summation over $\lambda$ is restricted to positive eigenvalues, $\Omega_\lambda>0$.
The transition matrix element between the QRPA ground state $|0\rangle$ and the excited state $|\lambda\rangle$ is 
\begin{subequations}
\begin{align}
 \langle \lambda | \hat{F}| 0\rangle &=  X^{\lambda\ast} \cdot F^{20}
 + Y^{\lambda\ast}\cdot F^{02}, \\
 \langle 0 | \hat{F}| \lambda\rangle &=  X^{\lambda} \cdot F^{02}
 + Y^{\lambda}\cdot F^{20} \,,
\end{align}
\end{subequations}
where we have used dots to signify the inner product of two-quasiparticle vectors, e.g.:  
\begin{align}
 X^{\lambda}\cdot F^{20} = \sum_{\mu<\nu} X^{\lambda}_{\mu\nu} F^{20}_{\mu\nu} \,.
\end{align}
The FAM response function is
\begin{align}
 S(\omega) &= 
 F^{20\ast}\cdot X(\omega) + F^{02^\ast}\cdot Y(\omega) \nonumber \\
 &= - \sum_{\lambda}
 \left[
 \frac{|\langle \lambda|\hat{F}|0\rangle|^2}{\Omega_\lambda - \omega}
 +
 \frac{ |\langle 0|\hat{F}|\lambda\rangle|^2}{\Omega_\lambda + \omega}
 \right] \,. \label{eq:strength}
\end{align}
The imaginary part of this function gives the strength distribution when imaginary-energy poles are absent,
\begin{align}
    \frac{dB}{d\omega}(\omega) = -\frac{1}{\pi}{\rm Im}\, S(\omega) \,.
\end{align}

\subsection{RBM} \label{sec:RBM}
In previous applications, the FAM equations (\ref{eq:SimpFAM}) had to be solved many times for different values of $\omega$.
We take a different approach, constructing an RBM-based FAM emulator to make fast predictions by treating the complex energy $\omega$ as an emulation parameter.
We start by carrying out FAM calculations for a small number of complex energies $\omega_j$ $(j=1, \cdots, n)$, values we call training energies.
Then we express the FAM amplitudes for an arbitrary value of $\omega$
as a superposition of the amplitudes at the training energies,
\begin{align}
\begin{pmatrix}
{\sf X}(\omega) \\ {\sf Y}(\omega)
\end{pmatrix}
&=
\sum_{j=1}^{n} \biggr\{
a_j(\omega)
\begin{pmatrix}
X(\omega_j) \\ Y(\omega_j)
\end{pmatrix}
+b_j(\omega)
\begin{pmatrix}
[Y(\omega_j)]^\ast \\ [X(\omega_j)]^\ast
\end{pmatrix} \nonumber \\
&\quad 
+a_{j+n}(\omega)
\begin{pmatrix}
[X(\omega_j)]^\ast \\ [Y(\omega_j)]^\ast
\end{pmatrix}
+b_{j+n}(\omega)
\begin{pmatrix}
Y(\omega_j) \\ X(\omega_j)
\end{pmatrix} \biggr\} \nonumber \\
&=
\sum_{j=1}^{2n} \biggr\{
a_j(\omega)
\begin{pmatrix}
X(\omega_j) \\ Y(\omega_j)
\end{pmatrix}
+b_j(\omega)
\begin{pmatrix}
[Y(\omega_j)]^\ast \\ [X(\omega_j)]^\ast
\end{pmatrix} \biggr\} \,. \label{eq:emulatorFAMamplitude}
\end{align}
Here, for a given complex training energy $\omega_j$, we include four basis states. 
From Eq.~(\ref{eq:FAMXY}), $X_{\mu\nu}(\hat{F}^\dag, -\omega^\ast) = [Y_{\mu\nu}(\hat{F},\omega)]^\ast$ and, therefore, the basis states weighted with $b_j(\omega)$ ($j=1, \cdots, n)$ correspond to the FAM amplitudes at $-\omega_j^\ast$ for the Hermitian conjugate operator $\hat{F}^\dag$. 
Similarly, the basis states weighted with $a_{j+n}(\omega)$ and $b_{j+n}(\omega)$ correspond to the FAM amplitudes at $\omega_j^\ast$ and $-\omega_j$.
The relations  $X_{\mu\nu}(\hat{F},\omega^\ast)=[X_{\mu\nu}(\hat{F}^\dag, \omega)]^\ast$  and $Y_{\mu\nu}(\hat{F}, \omega^\ast)=[Y_{\mu\nu}(\hat{F}^\dag,\omega)]^\ast$  hold if all the QRPA eigenvectors are real.
Hereafter, the index $j$ of the training energies runs from 1 to $2n$,  and the corresponding FAM amplitudes from $j=n+1$ to $2n$ are given by the complex conjugate of the FAM amplitudes from $j=1$ to $n$, i.e., 
$X(\omega_{n+j}) \equiv [X(\omega_j)]^\ast$ and  $Y(\omega_{n+j}) \equiv [Y(\omega_j)]^\ast$. 
The emulator is applicable even if $\hat{F}$ is not Hermitian and/or some of the QRPA eigenvectors are not real, because the FAM amplitudes serve just as basis vectors, and the weights  $a_j(\omega)$ and $b_j(\omega)$ are determined by the variational principle in the next step.

To proceed further, we impose the condition that the emulator FAM amplitudes in Eq.\ (\ref{eq:emulatorFAMamplitude}) satisfy the 
linear-response equation, Eq.\ (\ref{eq:FAMeq}). 
By multiplying those amplitudes by a matrix at the training energy $\omega_i$ $(i=1,\cdots, 2n)$ from the left, we have
\begin{align}
{\cal X}^\dag(\omega_i)
 [ {\cal S} - \omega \Sigma_3]
 \begin{pmatrix} {\sf X}(\omega) \\ {\sf Y}(\omega) \end{pmatrix}
 = 
 - \begin{pmatrix} {\sf S}^\ast_i \\ {\sf S}'_i  \end{pmatrix} \,,
 \label{eq:emulatorFAM}
\end{align}
where 
\begin{align}
{\cal X}^\dag(\omega_i) \equiv  \begin{pmatrix} [X(\omega_i)]^\ast & [Y(\omega_i)]^\ast \\ Y(\omega_i) & X(\omega_i) \end{pmatrix}
\end{align}
is a matrix composed of the FAM amplitudes at a training energy $\omega_i$, and the $2n$-dimensional vectors ${\sf S}_i $ and ${\sf S}_i'$ are given by
\begin{align}
{\sf S}_i = S(\hat{F},\omega_i), \quad
 {\sf S}_i'= S(\hat{F}^\dag,\omega_i) \quad (i=1,\cdots,2n) \,.
\end{align}
The left-hand side of Eq.~(\ref{eq:emulatorFAM}) contains the 
matrix elements of two Hermitian matrices, constructed from 
the QRPA norm kernel and the Hamiltonian sandwiched between two FAM amplitudes,
\begin{align}
{\sf N}_{ij} &\equiv {\cal X}^\dag(\omega_i) \Sigma_3 {\cal X}(\omega_j)
= \begin{pmatrix}
    {\sf N}^{(1)}_{ij} & {\sf N}^{(2)}_{ij} \\ -{\sf N}^{(2)\ast}_{ij}  & -{\sf N}^{(1)\ast}_{ij}
\end{pmatrix} \label{eq:normkernel}
\,, \\
{\sf H}_{ij} &\equiv {\cal X}^\dag(\omega_i){\cal S} {\cal X}(\omega_j)
= \begin{pmatrix}
    {\sf A}_{ij} & {\sf B}_{ij} \\ {\sf B}_{ij}^\ast  & {\sf A}_{ij}^\ast
\end{pmatrix} \,.
\end{align}
Their matrix elements are given by
\begin{subequations}
\begin{align}
{\sf N}^{(1)}_{ij} &= 
[X(\omega_i)]^\ast \cdot X(\omega_j) - [Y(\omega_i)]^\ast \cdot Y(\omega_j) \,, \\
{\sf N}^{(2)}_{ij} &= 
[X(\omega_i)]^\ast \cdot [Y(\omega_j)]^\ast - [Y(\omega_i)]^\ast \cdot [X(\omega_j)]^\ast \,,
\end{align}
\end{subequations}
and
\begin{subequations}
\begin{align}
{\sf A}_{ij} &= 
    [X(\omega_i)]^\ast\cdot [
    A X(\omega_j) + 
    B Y(\omega_j)
    ] \nonumber \\ & \quad
    +
    [Y(\omega_i)]^\ast\cdot [
    B^\ast X(\omega_j) + 
    A^\ast Y(\omega_j)
    ] \nonumber \\
    &=
    [X(\omega_i)]^\ast\cdot
    \delta H^{20}(\omega_j) 
    +
    [Y(\omega_i)]^\ast\cdot
    \delta H^{02}(\omega_j)\nonumber \\ & \quad
    + 
      [X(\omega_i)]^\ast\cdot \overline{E} X(\omega_j) 
    + [Y(\omega_i)]^\ast\cdot \overline{E} Y(\omega_j), \\
    {\sf B}_{ij} &=
    [X(\omega_i)]^\ast\cdot \{
    A [Y(\omega_j)]^\ast + 
    B [X(\omega_j)]^\ast
    \}\nonumber \\ & \quad
    +
    [Y(\omega_i)]^\ast\cdot \{
    B^\ast [Y(\omega_j)]^\ast + 
    A^\ast [X(\omega_j)]^\ast
    \} \nonumber \\
    &=
    [X(\omega_i)]^\ast \cdot
    [\delta H^{20}(\omega_j)]^\ast
    +
    [Y(\omega_i)]^\ast\cdot
    [\delta H^{02}(\omega_j)]^\ast\nonumber \\ & \quad
    + 
      [X(\omega_i)]^\ast \cdot \overline{E} [Y(\omega_j)]^\ast
    + [Y(\omega_i)]^\ast \cdot \overline{E} [X(\omega_j)]^\ast \,,
\end{align}
\end{subequations}
where 
\begin{align}
 \overline{E}_{\mu\nu,\mu'\nu'} = (E_\mu + E_\nu) \delta_{\mu\mu'} \delta_{\nu\nu'} \,.
\end{align}
One can avoid calculating the $A$ and $B$ matrices by using the induced fields
$\delta H^{20}$ and $\delta H^{02}$ at the training energies.

Next, we write Eq.\ (\ref{eq:emulatorFAM}) as
\begin{subequations}
\begin{align}
\sum_{j=1}^{2n} ( {\sf H}_{ij} - \omega {\sf N}_{ij})
 a_j(\omega) &= -  {\sf S}_i^\ast\,, \\
\sum_{j=1}^{2n} ( {\sf H}_{ij} - \omega {\sf N}_{ij})
 b_j(\omega) &= - 
 {\sf S}_i'\,. 
\end{align} \label{eq:ROeq}
\end{subequations}
This is a $4n$-dimensional set of reduced-order linear-response equations, which the FAM emulator amplitudes in Eq.\ (\ref{eq:emulatorFAMamplitude}) should satisfy.
The coefficients $a_j(\omega)$ and $b_j(\omega)$ can be obtained by solving these equations for a given value of $\omega$, and the emulator provides the FAM amplitudes for arbitrary $\omega$. 
The matrix elements of ${\sf H}$ and ${\sf N}$ do not depend on $\omega$, and finding solutions for arbitrary values of $\omega$ does not require any high-fidelity calculations with all the two-quasiparticle states.

To avoid numerical instability associated with the overcompleteness, however, we proceed by solving the generalized eigenvalue problem and subsequently constructing the matrix $[{\sf H}-\omega{\sf N}]^{-1}$.
We then obtain the emulator's strength function  by substituting Eqs.~(\ref{eq:emulatorFAMamplitude}) and (\ref{eq:ROeq}) into Eq.~(\ref{eq:strength}):
\begin{align}
S(\hat{F},\omega) 
&= \sum_{i=1}^{2n} a_i(\omega) {\sf S}_i
+ b_i(\omega) {\sf S}_i^{\prime\ast} \nonumber \\
&= -{\sf S} [{\sf H} - \omega{\sf N}]^{-1} {\sf S}^\ast 
   -{\sf S}^{\prime\ast} [{\sf H} - \omega{\sf N}]^{-1} {\sf S}' \,.
   \label{eq:emulatorstrength}
\end{align}
We thus see that $S(\hat{F},\omega)$ involves a sum over first-order poles, the
locations of which are given by the eigenvalues $\tilde{\Omega}_k$ in the equation 
\begin{align}
    {\sf H} f_\lambda = \tilde{\Omega}_\lambda {\sf N} f_\lambda \,, \label{eq:HillWheeler}
\end{align}
where the $f_\lambda$ are weight functions. 
The number of poles is equivalent to the dimension of the training subspace, $4n$. 
Equation~(\ref{eq:HillWheeler}) is equivalent to the Hill-Wheeler equation in the generator coordinate method (GCM) \cite{Ring-Schuck}.
The latter can be derived from a variational principle when many-body states are 
expressed as superpositions of non-orthogonal generalized Slater determinants.
The emulator can be regarded as an application of the GCM to the QRPA, with the FAM amplitude as basis states and the complex energy as the generator coordinate.
The matrices ${\sf N}$ and ${\sf H}$ correspond to the norm kernel and Hamiltonian kernel, 
which contain the overlaps of the basis Slater determinants and the matrix elements of the Hamiltonian operator among them. 
We solve Eq.~(\ref{eq:HillWheeler}) in the conventional manner used for the Hill-Wheeler equation in the GCM.
We first diagonalize the norm kernel,
\begin{align}
{\sf N} = {\sf U} \bar{{\sf N}} {\sf U}^{\dag} \,,
\end{align}
by using a unitary matrix ${\sf U}$ and a diagonal matrix $\bar{\sf N}$ of the form
\begin{align}
\label{eq:normdiag}
 {\sf U} = \begin{pmatrix} {\sf U}^{(1)} & {\sf U}^{(2)\ast} \\ {\sf U}^{(2)} & {\sf U}^{(1)\ast} \end{pmatrix}\,, \quad
 \bar{\sf N} = \begin{pmatrix} {\sf n} & 0 \\ 0 & -{\sf n} \end{pmatrix} \,,
\end{align}
where ${\sf n}_{ij} = \delta_{ij} n_i$, with eigenvalues $n_i>0$.
Unlike in the conventional GCM, where all eigenvalues are positive, half of the eigenvalues in Eq.\ (\ref{eq:normdiag}) are negative. 

The square root of the norm kernel can be defined by 
\begin{align}
{\sf N}^{\frac{1}{2}} = {\sf U} \bar{{\sf N}}^{\frac{1}{2}}  {\sf U}^{\dag}\,,
\end{align}
and when computing its inverse matrix ${\sf N}^{-\frac{1}{2}}$, we introduce a cutoff for the norm kernel,
restricting its eigenvalues to those with $|n_i|>n_{\rm cut}$ to avoid numerical instability.
We choose the following phases: 
\begin{align}
 \bar{\sf N}^{-\frac{1}{2}} = \begin{pmatrix} {\sf n}^{-\frac{1}{2}} & 0 \\ 0 &  -i {\sf n}^{-\frac{1}{2}} \end{pmatrix}
\end{align}
with ${\sf n}^{-\frac{1}{2}}_{ij} = \delta_{ij} n_i^{-\frac{1}{2}}$.
We can then write the denominator in the strength function as
\begin{align}
{\sf H} - \omega {\sf N}
= {\sf N}^{\frac{1}{2}}{\sf U} [ {\sf H}_{\rm coll} - \omega] {\sf U}^{\dag} {\sf N}^{\frac{1}{2}}\,,
\end{align}
where the collective Hamiltonian ${\sf H}_{\rm coll}$, which represents the action of the Hamiltonian in the collective subspace, is defined by 
\begin{subequations}
\begin{align}
    {\sf H}_{\rm coll} 
 &=  {\sf U}^{\dag} {\sf N}^{-\frac{1}{2}}{\sf H}{\sf N}^{-\frac{1}{2}} {\sf U}
 = \bar{\sf N}^{-\frac{1}{2}} {\sf U}^{\dag} {\sf H}{\sf U} \bar{\sf N}^{-\frac{1}{2}}
 \label{eq:collH} \nonumber \\ &
 = \begin{pmatrix} 
    {\sf A}_{\rm coll} & {\sf B}_{\rm coll} \\ -{\sf B}^\ast_{\rm coll} & - {\sf A}^\ast_{\rm coll}
 \end{pmatrix},\\
 {\sf A}_{\rm coll} &=
 n^{-\frac{1}{2}}(
{\sf U}^{(1)\dag} {\sf A}{\sf U}^{(1)}+
{\sf U}^{(1)\dag} {\sf B}{\sf U}^{(2)} \nonumber \\ &\quad +
{\sf U}^{(2)\dag} {\sf B}^\ast{\sf U}^{(1)}+
{\sf U}^{(2)\dag} {\sf A}^\ast{\sf U}^{(2)}
 ) n^{-\frac{1}{2}},\\
 {\sf B}_{\rm coll} &= \frac{1}{i}
 n^{-\frac{1}{2}}(
{\sf U}^{(1)\dag} {\sf A}{\sf U}^{(2)\ast}+
{\sf U}^{(1)\dag} {\sf B}{\sf U}^{(1)\ast} \nonumber \\ &\quad +
{\sf U}^{(2)\dag} {\sf B}^\ast{\sf U}^{(2)\ast}+
{\sf U}^{(2)\dag} {\sf A}^\ast{\sf U}^{(1)\ast}
 ) n^{-\frac{1}{2}}\,,
\end{align}
\end{subequations}
The collective Hamiltonian has the same symmetries as does the QRPA Hamiltonian, i.e.\ ${\sf A}_{\rm coll}$ is Hermitian and ${\sf B}_{\rm coll}$ is symmetric.
The eigenvectors and eigenvalues therefore do as well. Here we assume that the eigenvalues of the collective Hamiltonian are real for simplicity, 
and postpone the discussion of the more general case with imaginary eigenvalues to Appendix~\ref{sec:PQrepresentation}.

We note that the reduced-order eigenvalue problem in Eq.~\eqref{eq:HillWheeler}, or equivalently the collective Hamiltonian, is formally similar to the small parametric matrices employed in the PMM~\cite{Cook:2024toj}. In the PMM, however, the matrix elements are trainable parameters fixed by minimizing a cost function against training data. Here they are computed exactly from the FAM amplitudes at the training energies, and the eigenvalues $\tilde{\Omega}_\lambda$ are variational approximations to the QRPA energies.

The collective Hamiltonian can be written in the form
\begin{align}
{\sf H}_{\rm coll} = {\sf G} \tilde{\sf O} {\sf G}^{-1} \label{eq:diagHcoll} \,,
\end{align}
with the diagonal matrix $\tilde{\sf O}$
and the regular matrix ${\sf G}$ given by 
\begin{align}
 \tilde{\sf O} = \begin{pmatrix} \tilde{\sf \Omega} & 0  \\ 0  & -\tilde{\sf \Omega} \end{pmatrix}, \quad
 {\sf G} = \begin{pmatrix} {\sf G}^{(1)} & {\sf G}^{(2)\ast}  \\ {\sf G}^{(2)}  & {\sf G}^{(1)\ast}  \end{pmatrix} \,.
 \label{eq:G}
\end{align}
Here $\tilde{\sf \Omega}_{\lambda\mu} = \delta_{\lambda\mu} \tilde{\Omega}_\lambda$, and
the positive eigenvalues of the collective Hamiltonian, $\tilde{\Omega}_i>0$, provide the QRPA eigenvalues in the reduced-order space.
We require that the regular matrix ${\sf G}$ be normalized under the same condition as the amplitudes in Eq.~(\ref{eq:XYnormalization}),
${\sf G}^\dag\Sigma_3{\sf G} = \Sigma_3$ and ${\sf G}\Sigma_3{\sf G}^\dag = \Sigma_3$.  We then have 
\begin{align}
{\sf G}^{-1} &= \Sigma_3 {\sf G}^\dag\Sigma_3 = 
\begin{pmatrix}
 {\sf G}^{(1)\dag} & -{\sf G}^{(2)\dag} \\
 -{\sf G}^{(2)T} & {\sf G}^{(1)T} 
\end{pmatrix}. \label{eq:Ginv}
\end{align}
In terms of the matrices 
\begin{subequations}
\begin{align}
 {\sf U}\bar{\sf N}^{-\frac{1}{2}} {\sf G} &= 
 \begin{pmatrix}
 \tilde{\sf U} & -i \tilde{\sf V}^{\ast} \\ \tilde{\sf V} & -i \tilde{\sf U}^{\ast}
 \end{pmatrix}, \\
 {\sf G}^{-1}\bar{\sf N}^{-\frac{1}{2}} {\sf U}^\dag &= 
 \begin{pmatrix}
 \tilde{\sf U}^{\dag} & \tilde{\sf V}^{\dag} \\ -i\tilde{\sf V}^{T} & -i \tilde{\sf U}^{T}
 \end{pmatrix} \,,
 \end{align}
 \label{eq:ung}
\end{subequations}
with the matrix elements $\tilde{\sf U}$ and $\tilde{\sf V}$ given by
\begin{subequations}
\begin{align}
 \tilde{\sf U} &= {\sf U}^{(1)} {\sf n}^{-\frac{1}{2}} {\sf G}^{(1)} -i {\sf U}^{(2)\ast} {\sf n}^{-\frac{1}{2}} {\sf G}^{(2)}, \\
 \tilde{\sf V} &= {\sf U}^{(2)} {\sf n}^{-\frac{1}{2}} {\sf G}^{(1)} -i {\sf U}^{(1)\ast} {\sf n}^{-\frac{1}{2}} {\sf G}^{(2)}\,,
\end{align}
\end{subequations}
the response function and the coefficients of the FAM amplitudes in the RBM are
\begin{widetext}
\begin{align}
 [{\sf H} - \omega  {\sf N}]^{-1}
&={\sf U}\bar{\sf N}^{-\frac{1}{2}}{\sf G} [\tilde{\sf O} - \omega]^{-1}  {\sf G}^{-1}
 \bar{\sf N}^{-\frac{1}{2}} {\sf U}^\dag \nonumber \\ 
 &= 
 \begin{pmatrix}
    \tilde{\sf U}(\tilde{\sf \Omega}-\omega)^{-1}\tilde{\sf U}^{\dag} + \tilde{\sf V}^{\ast}(\tilde{\sf \Omega} + \omega)^{-1}\tilde{\sf V}^{T} & 
    \tilde{\sf U}(\tilde{\sf \Omega}-\omega)^{-1}\tilde{\sf V}^{\dag} + \tilde{\sf V}^{\ast}(\tilde{\sf \Omega} + \omega)^{-1}\tilde{\sf U}^{T} \\
    \tilde{\sf V}(\tilde{\sf \Omega}-\omega)^{-1}\tilde{\sf U}^{\dag} + \tilde{\sf U}^{\ast}(\tilde{\sf \Omega} + \omega)^{-1}\tilde{\sf V}^{T} & 
    \tilde{\sf V}(\tilde{\sf \Omega}-\omega)^{-1}\tilde{\sf V}^{\dag} + \tilde{\sf U}^{\ast}(\tilde{\sf \Omega} + \omega)^{-1}\tilde{\sf U}^{T}
 \end{pmatrix}, \label{eq:H-wNinv}
\end{align}
\begin{align}
\begin{pmatrix} a(\omega) \\ b(\omega) \end{pmatrix} =
- \begin{pmatrix} 
\tilde{\sf U}        (\tilde{\sf \Omega} - \omega)^{-1} \tilde{\sf S}^{(1)\ast} +
\tilde{\sf V}^{\ast} (\tilde{\sf \Omega} + \omega)^{-1} \tilde{\sf S}^{(2)\ast} \\
\tilde{\sf V}        (\tilde{\sf \Omega} - \omega)^{-1} \tilde{\sf S}^{(1)\ast} + 
\tilde{\sf U}^{\ast} (\tilde{\sf \Omega} + \omega)^{-1} \tilde{\sf S}^{(2)\ast} 
\end{pmatrix}  \,, \label{eq:emulatorab}
\end{align}
\end{widetext}
with
\begin{subequations}
\begin{align}
\tilde{\sf S}^{(1)} &= \tilde{\sf U}^{T} {\sf S} + \tilde{\sf V}^{T} {\sf S}^{\prime\ast}, \\ 
\tilde{\sf S}^{(2)} &= \tilde{\sf V}^{\dag} {\sf S} + \tilde{\sf U}^{\dag} {\sf S}^{\prime\ast} \,.
\end{align}
\end{subequations}
From Eqs.~(\ref{eq:emulatorstrength}) and (\ref{eq:emulatorab}),
the emulated strength function is given by
\begin{align}
S(\hat{F},\omega) &= - \tilde{\sf S}^{(1)} (\tilde{\sf \Omega} - \omega)^{-1} \tilde{\sf S}^{(1)\ast}
- \tilde{\sf S}^{(2)} ( \tilde{\sf \Omega} + \omega)^{-1} \tilde{\sf S}^{(2)\ast}
\nonumber \\ &= -\sum_{\lambda}
    \left[
    \frac{ |\langle \tilde{\lambda}|\hat{F}|0\rangle|^2}{\tilde{\Omega}_\lambda - \omega}
    + 
    \frac{ |\langle 0|\hat{F}|\tilde{\lambda}\rangle|^2}{\tilde{\Omega}_\lambda + \omega} \right] \,.
\end{align}
The squared transition strengths in this equation are given by
\begin{subequations}
\begin{align}
    |\langle \tilde{\lambda}|\hat{F}|0\rangle|^2 &\equiv 
    |\tilde{\sf S}^{(1)}_\lambda|^2, \\
    |\langle 0|\hat{F}|\tilde{\lambda}\rangle|^2 &\equiv 
    |\tilde{\sf S}^{(2)}_\lambda|^2 \,.
\end{align}
\end{subequations}
From Eqs.~(\ref{eq:emulatorFAMamplitude}) and (\ref{eq:emulatorab}),
the FAM emulator amplitudes can be written explicitly in the form
\begin{subequations}
\begin{align}
 {\sf X}(\omega) &= -\tilde{\sf X}(\tilde{\sf \Omega} - \omega)^{-1} \tilde{\sf S}^{(1)\ast}
 - \tilde{\sf Y}^{\ast} (\tilde{\sf \Omega} + \omega)^{-1}\tilde{\sf S}^{(2)\ast}, \\
 {\sf Y}(\omega) &= -\tilde{\sf Y}(\tilde{\sf \Omega} - \omega)^{-1} \tilde{\sf S}^{(1)\ast}
 - \tilde{\sf X}^{\ast} (\tilde{\sf \Omega} + \omega)^{-1}\tilde{\sf S}^{(2)\ast}\,,
\end{align}
\label{eq:emulatorFAMamplitude2}
\end{subequations}
where 
\begin{subequations}
\begin{align}
 \tilde{\sf X}^{\lambda}_{\mu\nu} &= \sum_{i=1}^{2n}X_{\mu\nu}(\omega_i) \tilde{\sf U}_{i\lambda} + Y^\ast_{\mu\nu}(\omega_i) \tilde{\sf V}_{i\lambda},\label{eq:Emulator_QRPA_X}\\
 \tilde{\sf Y}^{\lambda}_{\mu\nu} &= \sum_{i=1}^{2n}Y_{\mu\nu}(\omega_i) \tilde{\sf U}_{i\lambda} + X^\ast_{\mu\nu}(\omega_i) \tilde{\sf V}_{i\lambda} \,,
 \label{eq:Emulator_QRPA_Y}
\end{align}
\label{eq:emulator_QRPA_XY}
\end{subequations}
are the reduced-order QRPA eigenvectors.

The FAM emulator is intended to provide the FAM amplitudes at an arbitrary complex energy, and does so through Eq.~(\ref{eq:emulatorFAMamplitude2}).
However, Eqs.~(\ref{eq:diagHcoll}) and (\ref{eq:emulator_QRPA_XY}) show that it also emulates the QRPA eigenvalues and eigenvectors; the second set of equations expresses the QRPA eigenvectors as a superposition of the FAM amplitudes at training energies. Our framework thus provides a reduced-order solution of the QRPA 
problem in the FAM-amplitude basis.

\section{benchmark results \label{sec:results}}

\subsection{\texorpdfstring{Giant resonance from the oblate HFB state in $^{24}$Mg}{Giant resonance from the oblate HFB state in 24Mg}}

We use the axial HFB code {\sc hfbtho} \cite{Stoitsov200543, Stoitsov20131592,PEREZ2017363,MAREVIC2022108367} and its extension to the FAM solver for an axially symmetric mode \cite{PhysRevC.84.041305} to demonstrate the performance of the FAM-RBM emulator.
Following previous FAM studies~\cite{PhysRevC.84.041305, PhysRevC.87.064309,PhysRevC.91.044323}, we base our benchmark on an oblate axially-symmetric HFB ground state for $^{24}$Mg. 
Although $^{24}$Mg is actually prolate, and the oblate solution is not stable against triaxial deformation,  the oblate deformation has finite pairing gaps both in the neutron and proton channels, and thus provides a good starting point for tests of computational methods.

We choose to work with the EDF SLy4, with volume pairing of strength $V_0=-125.2$ MeV fm$^3$ for both
neutrons and protons, and without the center-of-mass correction. 
We employ harmonic-oscillator single-particle orbitals with up to $N_{\rm sh}=5$ shells, a restriction that is equivalent to a 60-MeV single-particle-energy cutoff. 
To obtain the same results as does Ref.\ \cite{PhysRevC.84.041305}, we adopt the Gauss-quadrature points $N_{\rm GH}=N_{\rm GL}=N_{\rm Leg}=30$ and set the length-scale parameter in the computation of the direct Coulomb functional to $L=b^4$ fm, where $b$ is the harmonic-oscillator length parameter in units of fm.

Now we intend to construct an emulator that can reproduce the giant resonance strength function up to 50 MeV. 
We begin by choosing complex training energies with real parts that are one MeV apart between 0 and 49 MeV and a fixed imaginary part. 
We consider three emulators, with 
Im $\omega_i=1$ MeV (emulator 1),
Im $\omega_i=5$ MeV (emulator 5), 
and 
Im $\omega_i=10$ MeV (emulator 10).
For the external field, we use the isoscalar monopole operator,
\begin{align}
\hat{F}_{\rm ISM} = \frac{Z}{A} \int d\bm{r} r^2 \sum_{st} \hat{c}^\dag_{\bm{r}st} \hat{c}_{\bm{r}st}\,, \label{eq:ISM}
\end{align}
where $\hat{c}^\dag_{\bm{r}st}$ and $\hat{c}_{\bm{r}st}$ create and annihilate nucleons at position $\bm{r}$, with spin $s$ and isospin $t$.

\subsubsection{Norm cutoff dependence}

\begin{figure}[!t]
\includegraphics[width=80mm]{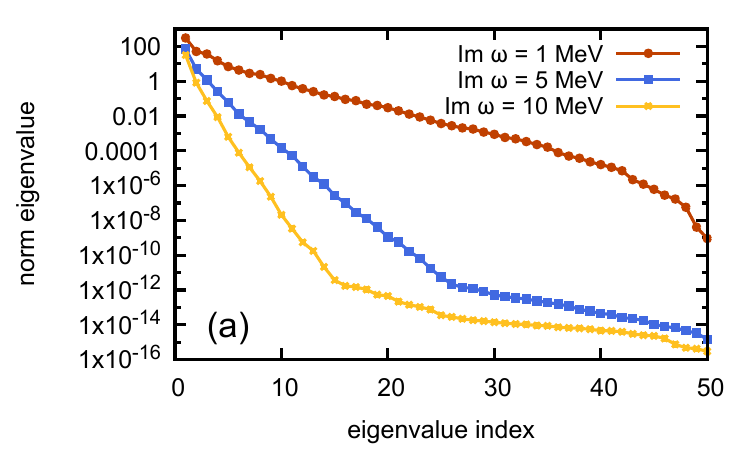} \\
\includegraphics[width=80mm]{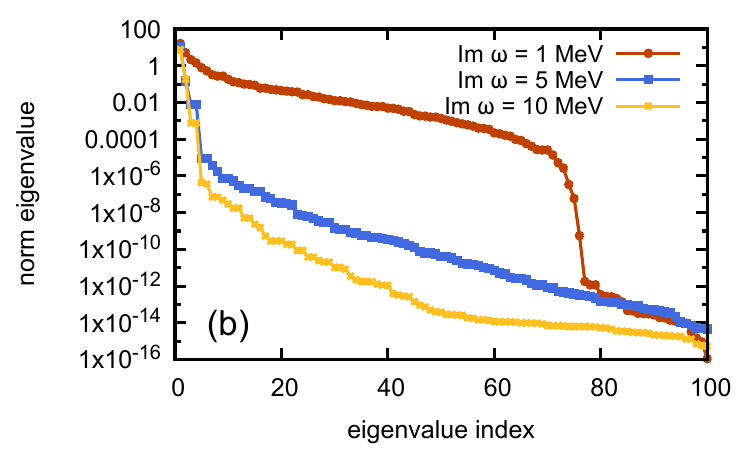}
\caption{Norm-kernel eigenvalues in descending order for (a) the 100-dimensional and (b) 200-dimensional emulators.
\label{fig:norm}}
\end{figure}

Figure~\ref{fig:norm} compares the eigenvalues of the norm-kernel matrix for three emulators.
The upper panel shows the eigenvalues for 100-dimensional ($n=50$) emulators, with FAM solutions at $\omega_i$ and $-\omega_i^\ast$ both contained in the training bases.
The lower panel shows the eigenvalues for the 200-dimensional emulators, which contain in addition FAM solutions at $-\omega_i$ and $\omega_i^\ast$.  
The 200-dimensional emulators have larger norm-kernel eigenvalues, indicating that they span a larger two-quasiparticle subspace. 
Emulators 5 and 10 have smaller norm eigenvalues 
than does emulator 1, for both the 100- and 200-dimensional cases, indicating that basis states with a smaller imaginary energy
span a larger subspace of the two-quasiparticle space.

In the following, we take four basis states for each complex training energy $\omega_i$ (as in the 200-dimensional emulators). The extra computing costs due to the additional solutions at $\omega_i^\ast$ and $-\omega_i$ are minimal, because no additional FAM calculations are required to obtain them.

\begin{figure}[!t]
\includegraphics[width=80mm]{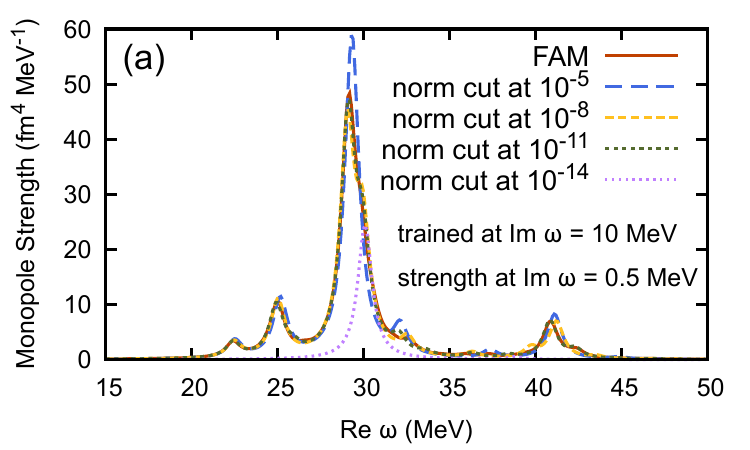} \\
\includegraphics[width=80mm]{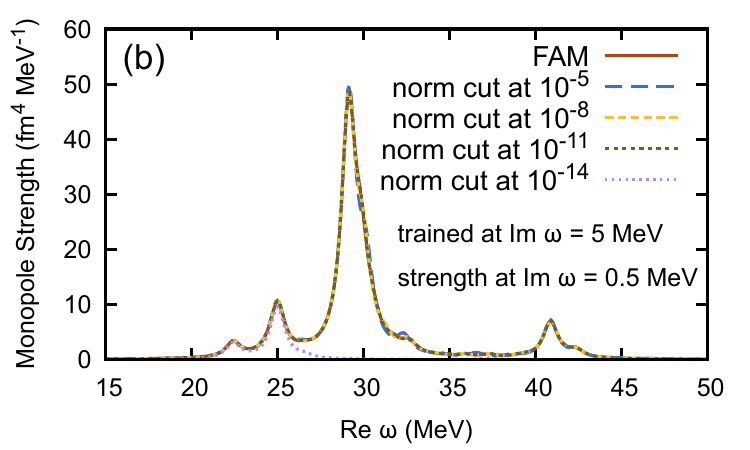} \\
\includegraphics[width=80mm]{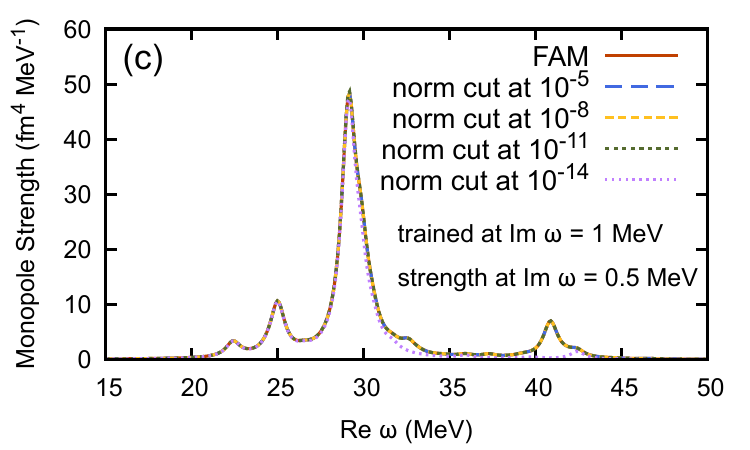} \\
\caption{Performance of emulators for the isoscalar monopole strength function in the giant resonance region. Dependence on the norm-eigenvalue cutoff for emulators (a) 10, (b) 5, and (c) 1. \label{fig:normdep}}
\end{figure}

Figure~\ref{fig:normdep} shows the isoscalar-monopole strength functions produced by the three emulators 10, 5, and 1 with ${\rm Im}\,\omega=0.5$ MeV alongside that of the full FAM calculation.
Each subfigure compares the results with four norm-eigenvalue cutoffs.
The strength function from emulator 10 shows a large dependence on the cutoff. 
The values $10^{-8}$ and $10^{-11}$ result in a fair approximation to the full FAM strength function, showing that the actual dimension included in this emulator is about 10--30 (see the yellow curve in the bottom panel of Fig.~\ref{fig:norm}) 
The strength functions calculated by the emulator 5 and 1 agree better, except for the calculation with a cutoff of $10^{-14}$, which may lead to large numerical error.

That the performance of the emulators depends on the imaginary part of the training energies can be understood as follows:
The FAM amplitudes evaluated at a training energy far from the real axis contain contributions from many QRPA poles, with the weight proportional to the strength, because the difference in the complex plane between the energy and those of the poles on the real axis varies less from pole to pole. 
(See Eq.~(\ref{eq:FAMXY}) for the relation between the FAM amplitudes and the QRPA eigenvectors.) 
The overlap between FAM-amplitude vectors at different training energies thus is often relatively large.
When the training energy is close to the real axis, the amplitudes are affected most by the nearby QRPA poles, and thus, the amplitude vectors at different training energies are closer to orthogonal and have significant components in a larger subspace.
 
\subsubsection{Dependence on training-energy interval}

Using emulator 1, which gives the best performance among the three, we investigate the dependence on the number of training-energy points.
We compare the 2-MeV-interval (25 points), 3-MeV-interval (16 points),  4-MeV-interval (12 points), and 5-MeV-interval (10 points) emulators.

\begin{figure}[!t]
\includegraphics[width=80mm]{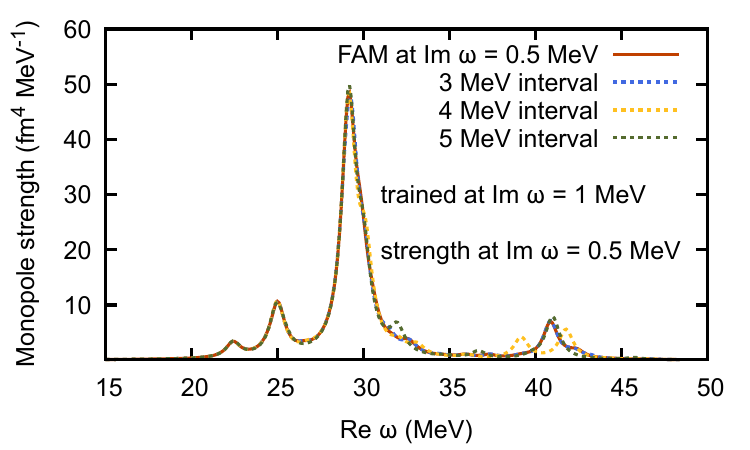}
\caption{The isoscalar monopole strength function calculated with the emulator 1, highlighting the dependence on the training-energy interval.
\label{fig:trainingpointdep}}
\end{figure}

Figure~\ref{fig:trainingpointdep} shows the resulting strength distributions.  We omit the results with 2-MeV intervals because they are not visibly different from those with 1-MeV intervals in Fig.\ \ref{fig:normdep}.
All the emulators correctly describe the giant resonance near 30 MeV.
A tiny deviation is visible around 32 MeV in the results with 3-MeV intervals.
The calculations with 4- and 5-MeV intervals do a poorer job at higher energies.
The number of training points necessary to reproduce the strength distribution may depend on the nucleus and excitation mode, but our calculation shows that a 1-MeV interval with ${\rm Im}\, \omega = 1$ MeV is more than sufficient to reproduce the giant-resonance strength distribution.

Recent studies \cite{2511.01844,2511.10420} have explored so-called greedy algorithms (an active learning approach), which choose training points within the parameter space in a step-by-step fashion. At each step, the algorithm identifies the new training point where the emulation errors, as estimated by the easy-to-calculate error proxies, are the largest. In this way, one can obtain a minimal set of training points, and thus have an adaptive method for deciding the effective dimension of the subspace, and with it the rank of the norm kernel.

\subsubsection{Strength function for other operators}

The FAM emulator provides approximate QRPA solutions. 
Thus, in principle, an emulator trained with the FAM amplitudes for operator $\hat{F}$ can produce the strength function for another operator $\hat{F}'$.
Here we examine the distributions associated with the isoscalar quadrupole and isovector monopole/quadrupole operators, defined by
\begin{align}
 \hat{F}_{\rm ISQ} &= \frac{Z}{A}  \sqrt{\frac{5}{16\pi}}
 \int d\bm{r} (2z^2 - x^2 -y^2)
 \sum_{st}  \hat{c}_{\bm{r}st}^\dag \hat{c}_{\bm{r}st}, \\
 \hat{F}_{\rm IVM} &= 
 \int d\bm{r} r^2
 \sum_{s} \left(
 \frac{Z}{A}\hat{c}_{\bm{r}sn}^\dag 
 \hat{c}_{\bm{r}sn}
 -\frac{N}{A}
 \hat{c}_{\bm{r}sp}^\dag 
 \hat{c}_{\bm{r}sp}\right), \\
 \hat{F}_{\rm IVQ} &= 
 \sqrt{\frac{5}{16\pi}}
 \int d\bm{r} (2z^2 - x^2-y^2)
  \nonumber \\ 
 &\quad\sum_{s} \left(
 \frac{Z}{A}
 \hat{c}_{\bm{r}sn}^\dag \hat{c}_{\bm{r}sn}
 - 
 \frac{N}{A}
 \hat{c}_{\bm{r}sp}^\dag \hat{c}_{\bm{r}sp}\right)\,.
\end{align}

\begin{figure}[!t]
\includegraphics[width=80mm]{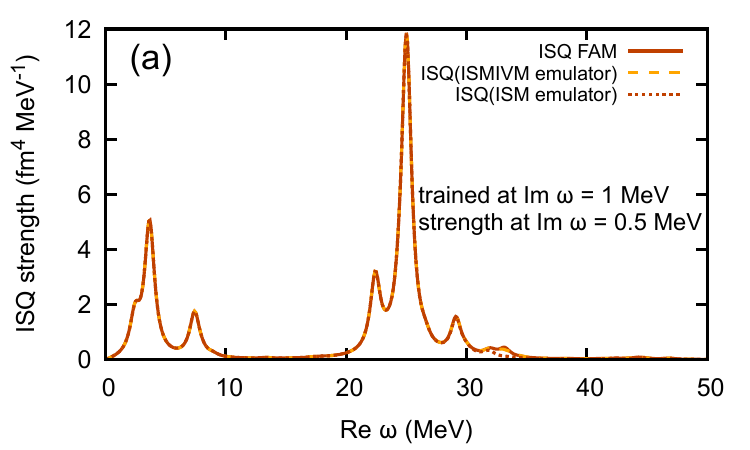} \\
\includegraphics[width=80mm]{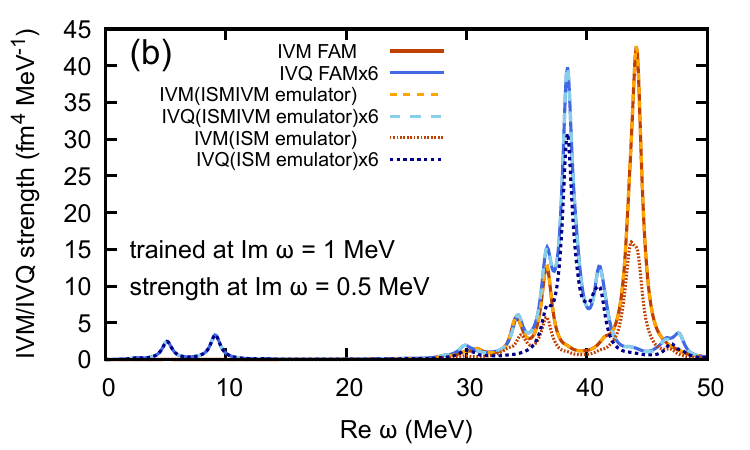} 
\caption{
(a) Performance of emulators trained with the isoscalar monopole operator (ISM) and both the isoscalar and isovector monopole operators (ISMIVM) for the 
isoscalar quadrupole (ISQ) strength function.
(b) Performance of emulators trained with the ISM and ISMIVM operators for the isovector monopole (IVM) and isovector quadrupole (IVQ) strength functions.
\label{fig:isqivmivq}
}
\end{figure}

Figure~\ref{fig:isqivmivq} shows, among other things, the strength functions for these operators produced by an emulator that is trained on the isoscalar monopole operator. 
The emulator perfectly reproduces the strength for the isoscalar quadrupole operator, while it underestimates the strengths for the isovector monopole and quadrupole operators.
The isoscalar and isovector modes are relatively well decoupled because of approximate isospin symmetry in $^{24}$Mg, which has $N=Z$ (only the Coulomb EDF breaks isospin symmetry explicitly).
The positions of the peaks are well reproduced, however, even when using only the isoscalar operator for training.

For a more comprehensive description, the FAM amplitudes produced by the isovector monopole operator can be included in the basis states in Eq.~(\ref{eq:emulatorFAMamplitude}) by extending the training parameters  $\omega_j$ to a set of $(\omega_j, \hat{F}_j)$ that includes more than one operator. In Fig.~\ref{fig:isqivmivq}, the strength function computed with the emulator trained at 50 training energies with the isoscalar monopole operator and 50 training energies with the isovector monopole operator is also shown.
Including the latter operator leads to a perfect reproduction of the strength distributions, both for the isoscalar quadrupole mode and 
the two isovector modes.
Ideally, one should include the operator one is trying to emulate in the training set. Nevertheless, the results just presented imply that the set of operators required for the training can be significantly reduced in cases where operators depend on a continuous variable such as momentum transfer that varies over a large range. 
This is a major advantage of the RBM emulator over the FAM itself.

\subsection{\texorpdfstring{Low-energy emulator at the oblate HFB state at $^{24}$Mg}{Low-energy emulator at the oblate HFB state at 24Mg}}

We turn now to QRPA states with $K^\pi=0^+$ below 10 MeV.
We train with FAM amplitudes from the isoscalar or isovector monopole operator at 11 training energies, with ${\rm Re}\,\,\omega_i= 0, 1, \cdots, 10$ MeV and ${\rm Im}\,\,\omega_i = 1$ MeV.
Table \ref{table:24Mglowenergy} lists the resulting QRPA energies and 
isoscalar and isovector monopole strengths, computed by contour integration and their differences from the emulator values.
The emulators reproduce the full QRPA energies accurately in the training-energy region, particularly for states with
large transition strengths.  The
RBM emulator is thus especially powerful in describing collective states.
The transition strengths are reproduced less accurately than the QRPA energies, but 
Fig.~\ref{fig:24Mg_ism} shows that the agreement is enough to reproduce the strength distribution well.
The emulator that includes the isovector monopole operator in the set that generates training basis states fails to reproduce one QRPA state at 7.96 MeV, but that state has an isovector monopole
transition strength from the ground state (1.567$\times 10^{-8}$ fm$^4$), that is three orders of magnitude smaller than those to neighboring states.
Such non-collective states are generally not important in applications.

In addition to the low-energy poles listed in Table~\ref{table:24Mglowenergy}, we obtain a pair of near-zero-energy modes, with $\Omega_i\sim \pm 10^{-3}i$ MeV.
They may correspond to the neutron and proton pairing rotations associated with the number-gauge symmetry breaking, but could also arise from numerical errors associated with the norm kernel cutoff.
We do not recommend using an emulator to treat zero-energy modes; they are easy to obtain in a single FAM calculation \cite{PhysRevC.92.034321}.

\begin{table*}[!t]
\caption{
Low-energy QRPA properties for $K^\pi=0^+$ excitations based on the oblate HFB state of $^{24}$Mg. 
QRPA energies $\Omega_i$ from contour integration in the FAM with the isoscalar monopole operator
and the difference from the emulator energies $({\rm Re}\,\tilde{\Omega}_i) - \Omega_i$ with the 
isoscalar and isovector monopole operators are listed in units of MeV.
The isoscalar and isovector monopole strengths calculated with the contour integration in the FAM
and the difference from the emulator strength are compared in the last four columns
in units of fm$^4$. The numbers in the parentheses denote powers of 10.
Emulators are constructed with only the isoscalar or isovector monopole operators in the training. 
States with norm kernel eigenvalues smaller than $10^{-11}$ are removed. 
\label{table:24Mglowenergy}}
\begin{ruledtabular}
\begin{tabular}{ccccccc}
FAM & RBM(ISM)$-$FAM & RBM(IVM)$-$FAM & FAM & RBM$-$FAM & FAM & RBM$-$FAM \\ 
 \multicolumn{3}{c}{$\Omega_i$ (MeV)} & 
 \multicolumn{2}{c}{$|\langle i|\hat{F}^{\rm ISM}|0\rangle|^2$ (fm$^4$)} &
  \multicolumn{2}{c}{$|\langle i|\hat{F}^{\rm IVM}|0\rangle|^2$ (fm$^4$)}  \\ \hline
1.3183361 & \,\,\,\,\,0.0004200    & \,\,\,\,\,0.0001188 & $5.777(-4)$ & $\,\,\,\,\,7.873(-5)$        & $1.547(-3)$ & $\,\,\,\,\,5.580(-6)$\\
1.3730998 & $-0.0001625$  & \,\,\,\,\,0.0007234& $1.510(-2)$ & $\,\,\,\,\,7.887(-5)$  & $5.822(-5)$ & $-5.493(-6)$\\
2.4582479 & \,\,\,\,\,0.0000142  & \,\,\,\,\,0.0026949 & $1.781(-1)$ & $\,\,\,\,\,8.641(-6)$  & $2.049(-6)$ & $\,\,\,\,\,4.252(-8)$\\
2.5975513 & \,\,\,\,\,0.0000022 & $-0.0000164$ & $3.062(-3)$ & $\,\,\,\,\,1.217(-5)$  & $8.904(-5)$ & $-8.230(-8)$\\
3.6670821 & $-0.0000001$ & $-0.0000430$ & $5.782(-1)$ & $-4.068(-6)$  & $1.584(-5)$ & $\,\,\,\,\,3.791(-8)$\\
5.1190160 & \,\,\,\,\,0.0000026 & $-0.0000001$ & $3.739(-4)$ & $\,\,\,\,\,1.182(-8)$  & $3.910(-2)$ & $-3.118(-7)$\\
7.4107121 & $-0.0000002$  & $-0.0000025$ & $4.879(-1)$ & $\,\,\,\,\,1.560(-6)$  & $1.442(-5)$ & $\,\,\,\,\,7.530(-8)$\\
7.8898871 & \,\,\,\,\,0.0002941  & \,\,\,\,\,0.0000001 & $8.756(-3)$ & $\,\,\,\,\,7.019(-5)$  & $2.850(-5)$ & $-1.735(-7)$\\
7.9604664 & \,\,\,\,\,0.0001106  &     --               & $3.295(-2)$ & $-6.935(-5)$  & $1.567(-8)$ & --\\
8.9285297 &  \,\,\,\,\,0.0000076 & $-0.0000168$ & $8.598(-2)$ & $-2.557(-6)$  & $2.975(-5)$ & $\,\,\,\,\,2.301(-6)$\\
9.1277704 & \,\,\,\,\,0.0003865 & \,\,\,\,\,0.0000247 & $2.342(-3)$ & $\,\,\,\,\,2.141(-6)$  & $1.177(-2)$ & $-2.476(-6)$\\
10.244543 & \,\,\,\,\,0.0017409  & \,\,\,\,\,0.0000060 & $2.246(-4)$ & $-4.090(-6)$  & $8.402(-4)$ & $-4.202(-8)$
\end{tabular}
\end{ruledtabular}
\end{table*}

\subsection{\texorpdfstring{Imaginary-energy solution from the spherical HFB state in $^{24}$Mg}{Imaginary-energy solution from the spherical HFB state in 24Mg}
}

We now consider the response of a spherical HFB state in $^{24}$Mg
with the same single-particle space and EDF as before.
This spherical state is not stable against quadrupole deformation, and a pair of imaginary-energy solutions appears in the axial quadrupole channel.

In Table~\ref{table:spherical24Mg}, we list energies and strengths obtained with FAM contour integration and the differences between the RBM and the FAM.
To evaluate the QRPA energies, we use the ratio of contour integrals
\begin{align}
 \Omega_i = \frac{\displaystyle \frac{1}{2\pi i}\oint_{C_i} \omega S(\hat{F},\omega) d\omega}
 {\displaystyle \frac{1}{2\pi i} \oint_{C_i} S(\hat{F},\omega) d\omega} \,,
\end{align}
where the contour $C_i$ includes one of the poles on the imaginary axis (we choose the one with ${\rm Im}\,\,\Omega_i>0$).
In the calculation of collective inertias for the five-dimensional collective Hamiltonian and for fission dynamics using local QRPA, imaginary-energy solutions may occur, requiring the use of the PQ representation
(see Appendix~\ref{sec:PQrepresentation} for a detailed discussion).
The transition strength in the PQ representation 
and the QRPA collective mass are obtained through the relations 
\begin{align}
 |\langle P_i|\hat{F}|0\rangle|^2 &=
 2\frac{1}{2\pi i} \oint_{C_i} \omega S(\hat{F},\omega) d\omega, \\
 {\cal M}_i &= |\langle P_i|\hat{F}|0\rangle|^{-2}\,.
\end{align}
See Ref.~\cite{PhysRevC.103.014306} for a detailed derivation.

We use the same training strategy to construct the RBM emulator, taking eleven complex-energy points at ${\rm Im}\,\,\omega_i = 1$ MeV, with 1 MeV intervals between $0\le{\rm Re}\,\,\omega_i\le 10$ MeV, 
for the isoscalar quadrupole operator.

In addition to the real low-energy eigenvalues and eigenvectors in the training region, the imaginary-energy eigenvalue and eigenvectors are precisely reproduced by the emulator.
We insist here that the training energies are chosen in the same way as in the previous examples, and the variational equation then automatically finds the imaginary-energy solutions. Moreover,
the transition strength and QRPA collective mass are precise, especially for the states with large transition strength (or equivalently, with smaller collective mass). The results show that the emulator is a powerful tool for quickly computing the local QRPA collective mass in large-amplitude collective motion, including fission and five-dimensional quadrupole dynamics, where one must compute the masses associated with a few collective operators from dozens of low-energy states, including the imaginary-energy solutions, and identify the most collective set of low-energy states.
\cite{PhysRevC.82.064313, PhysRevC.83.061302, PhysRevC.103.014306, PhysRevC.109.L051301}.

We do not discuss zero-energy modes associated with the symmetry breaking in this paper. 
Symmetries that become approximate because of numerical implementation 
such as the translational modes in a finite basis calculation, appear as spurious real or imaginary solutions in practical numerical calculations.
Zero-energy modes such as pairing rotation that are exact in numerical calculations 
give a discrete contribution to the FAM amplitudes and strength function at $\omega=0$ \cite{PhysRevC.92.034321}.
By avoiding use of the symmetry operator as an external field, however, one can completely separate the zero-energy modes from other physical modes.

\begin{table*}
\caption{
FAM results and RBM deviations for isoscalar quadrupole excitations in spherical $^{24}$Mg.
Shown are QRPA energies $\Omega_i$, the isoscalar transition strengths
$|\langle i | \hat{F}^{\rm ISQ}|0\rangle|^2$, the corresponding
strengths in the PQ representation $|\langle P_i|\hat{F}^{\rm ISQ}|0\rangle|^2$, and the QRPA masses ${\cal M}_i$ obtained with FAM contour integration, together with the 
corresponding differences between the RBM and FAM results 
(RBM$-$FAM) for states with absolute energies below 5 MeV.
The numbers in the parentheses denote powers of 10.
\label{table:spherical24Mg}}
\begin{ruledtabular}
\begin{tabular}{llllllll}
FAM & RBM$-$FAM & FAM & RBM$-$FAM & FAM & RBM$-$FAM & FAM & RBM$-$FAM \\ 
 \multicolumn{2}{c}{$\Omega_i$ (MeV)} & 
 \multicolumn{2}{c}{$|\langle i|\hat{F}^{\rm ISQ}|0\rangle|^2$ (fm$^4$)} &
  \multicolumn{2}{c}{$|\langle P_i|\hat{F}^{\rm ISQ}|0\rangle|^2$ (MeV fm$^4$)} &
 \multicolumn{2}{c}{${\cal M}_i$ (MeV$^{-1}$ fm$^{-4}$)} 
 \\ \hline
1.8596619$i$& \,\,\,\,\,0.0000006$i$ & \quad\quad--          &  \quad\quad--           & $7.629(+1)$ & $\,\,\,\,\,1.633(-5)$ & $1.311(-2)$ & $-2.806(-9)$ \\ \hline
1.7969748   & $-0.0000051$& $6.584(-4)$ & $\,\,\,\,\,5.375(-9)$ & $2.366(-3)$ & 
$\,\,\,\,\,1.084(-6)$ & $4.226(+2)$ & $-1.935(-1)$ \\
2.8946443   & $\,\,\,\,\,0.00000001$ & $5.029(0)$ & $-8.854(-7)$ & $2.911(+1)$ & 
$-2.491(-6)$& $3.435(-2)$ & $\,\,\,\,\,2.038(-9)$ \\
4.8579650   & $-0.0000074$ & $2.394(-4)$ & $-1.204(-8)$ & $2.326(-3)$ & 
$-9.851(-7)$& $4.299(+2)$ & $\,\,\,\,\,1.821(-1)$
\end{tabular}
\end{ruledtabular}
\end{table*}

\subsection{Comparison with iterative Arnoldi method}

The iterative Arnoldi method~\cite{PhysRevC.81.034312}
is an alternative scheme for solving the QRPA problem in a small subspace. 
The method generates an orthonormal Krylov subspace by repeatedly applying the QRPA matrix to a 
pivot vector and orthogonalizing the resulting vectors.
The method is based on the representation of the QRPA as linearized TDHFB,  
with the calculation of the $A$ and $B$ matrices avoided by computing
the FAM induced fields $\delta H^{20}$ and $\delta H^{02}$ from Arnoldi vectors that contain the FAM amplitudes. The detailed formulation is given in Appendix \ref{sec:Arnoldi}.

\begin{figure}[!t]
\includegraphics[width=80mm]{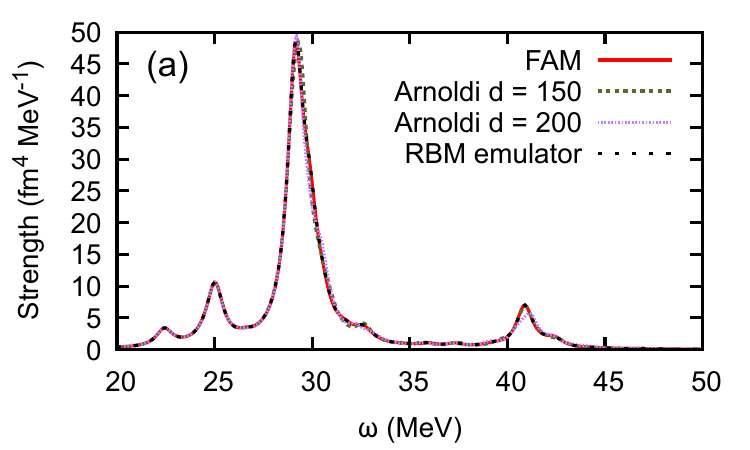} \\
\includegraphics[width=80mm]{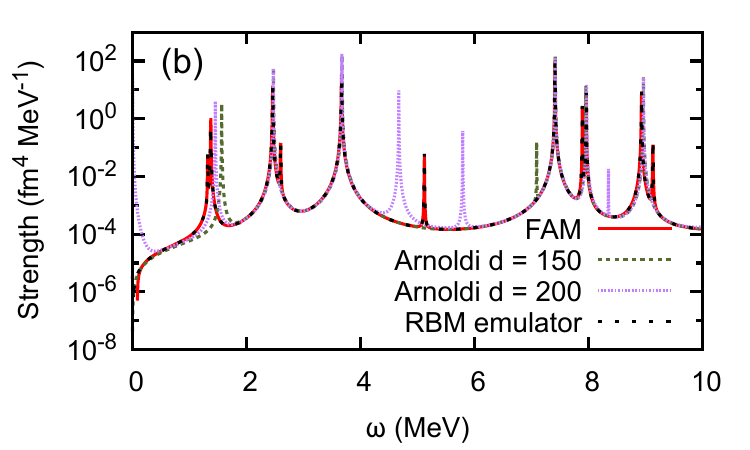} 
\caption{Isoscalar monopole strength distribution associated with the oblate $^{24}$Mg, calculated with the 
FAM, the iterative Arnoldi method with subspace dimensions $d=150$ and 200, and the RBM emulators trained at Im $\omega=1$ MeV, with
the norm-kernel cutoff set to $10^{-11}$.
(a) the giant resonance region and (b) the region below 10 MeV are shown. \label{fig:24Mg_ism}}
\end{figure}

In Fig.~\ref{fig:24Mg_ism},  we compare the performance of the 
iterative Arnoldi method to that of the RBM emulator for the isoscalar monopole strength distribution associated with the oblate $^{24}$Mg configuration. Although the iterative Arnoldi method with $d=150$ or 200 (similar dimensions to those of the RBM emulator) reproduces the exact FAM strength distribution in the giant resonance region, it misses or misaligns many peaks at low energies. Meanwhile, the RBM emulator can accurately reproduce the QRPA solutions in both energy regions, shown separately in the two panels; such excellent performance is expected because the emulator is trained in those regions. By contrast, the iterative Arnoldi method does not allow us to selectively improve the accuracy of QRPA solutions in a specific energy domain.
The comparison shows the advantage of the RBM emulator over the iterative Arnoldi method for low-energy QRPA states in, for example, the computation of collective inertia and low-energy excitations.

\section{\texorpdfstring{$\beta$}{beta} vibration \label{sec:beta}}

As realistic examples, we discuss the $K^\pi=0^+$ quadrupole vibrations in deformed Dy isotopes. We use the EDF SkM* with mixed-type pairing in a harmonic-oscillator single-particle model space containing up to $N_{\rm sh}=20$ shells.
We fit the pairing strengths to the odd-even mass-staggering parameters in $^{156}$Dy ($\Delta_n=1.175$ MeV and  $\Delta_p=0.976$ MeV), which are $V_n =-281.98$ MeV fm$^3$ and $V_p = -307.55$ MeV fm$^3$ with a 60 MeV equivalent-single-particle-energy cutoff. 
The calculated quadrupole deformation of $^{156}$Dy is $\beta_2=0.274$.

\subsection{Strength distribution}

We evaluate the isoscalar quadrupole strength distribution for 13 even-even Dy isotopes ($^{156-180}$Dy).
The computational time for the high-fidelity FAM calculations (iteration parts only)
of the strength distribution up to 25 MeV, with 0.1 MeV intervals (250 points) and ${\rm Im}\,\omega = 0.5$ MeV, is about 266.5 core hours for even-even $^{156-180}$Dy isotopes on the Miyabi-C supercomputer.

\begin{figure}[!t]
\includegraphics[width=80mm]{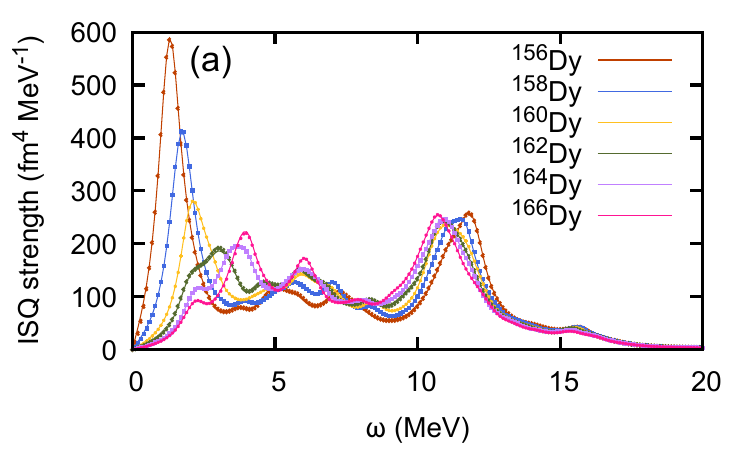} \\
\includegraphics[width=80mm]{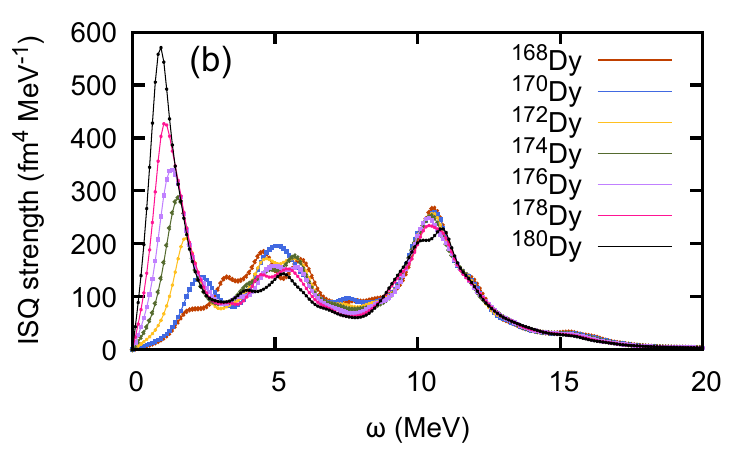}
\caption{Isoscalar quadrupole strength distributions for (a) the $^{156-166}$Dy and (b) $^{168-180}$Dy isotopes.
Dots are the strengths calculated in the FAM with an energy interval of 0.1 MeV.
Lines are the results of the emulator with 26 training points. Both are evaluated at ${\rm Im}\,\omega=0.5$ MeV. \label{fig:dy-gremulator}}
\end{figure}

We construct an emulator that covers the giant-resonance energy region by taking 26 training complex-energy points that consist of ${\rm Re}\,\omega=0, 1, 2, \cdots$, 25 MeV, with ${\rm Im}\,\omega=1.0$ MeV.
The total computational time for FAM iteration at these 26 points in $^{156-180}$Dy isotopes is 22.6 core hours, 
just 8.5\% of the conventional FAM calculations for the strength distribution functions.
Figure~\ref{fig:dy-gremulator} compares the resulting strength distributions with those of the full FAM at 0.1 MeV intervals. 
The emulator successfully describes the strength distribution across the whole energy region, including the low-energy and giant resonances, at a much lower computational cost. 
Discrepancies between the two calculations are negligible.

\subsection{Lowest-energy states}

We next focus on the lowest-energy excited states in the $K^\pi=0^+$ channel, corresponding to $\beta$ vibrations, at around 2 MeV in Fig.~\ref{fig:dy-gremulator}. 
With the full FAM, we use 500 points in the complex-energy plane,
from 0 to 5 MeV with 0.01 MeV intervals for ${\rm Re}\,\omega$ and with ${\rm Im}\,\omega=0.01$ MeV, to separate the lowest-energy poles.
The total computational time is about 682 core hours for 13 Dy isotopes.
We need an additional 18.4 core hours for the contour integration around the lowest-energy pole (discretized with 11 points at a distance of 0.02 MeV from the pole).

We construct FAM emulators for the low-energy solutions by taking 11 training points in the complex-energy plane, with ${\rm Re}\,\omega=0, 0.5, 1.0, \cdots, 5.0$ MeV, and with ${\rm Im}\,\omega=1.0$ MeV.
It takes a total of 10.1 core hours to train the emulator in all 13 isotopes.
It takes much less computational time to diagonalize the QRPA matrix in the RBM subspace spanned by the training points. The computational time for the FAM calculation 
to construct the RBM basis is just 1.4\% of the time for conventional contour integration.

\begin{figure}[!t]
\includegraphics[width=80mm]{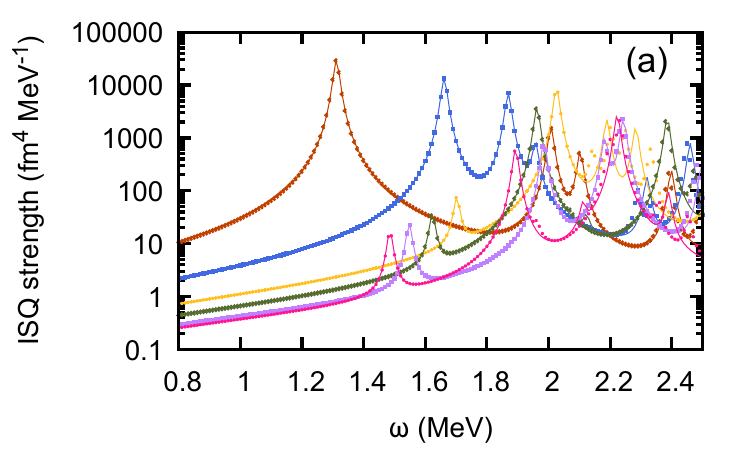} \\
\includegraphics[width=80mm]{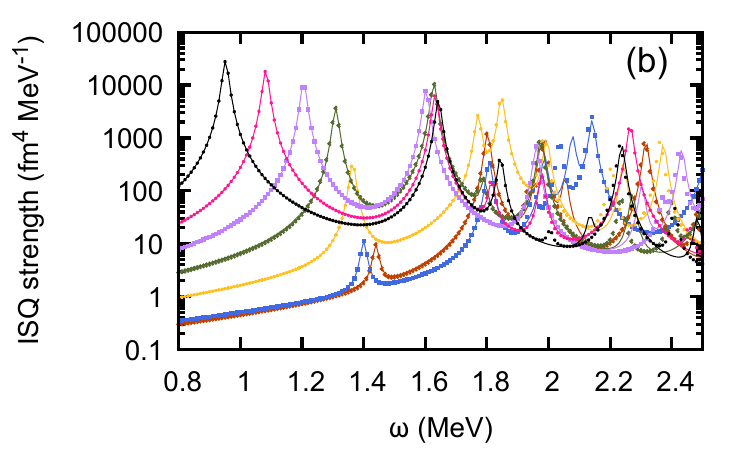}
\caption{Isoscalar quadrupole strength distribution for (a) the $^{156-166}$Dy and (b) $^{168-180}$Dy isotopes.
Dots are the strengths in the FAM with an energy interval of 0.01 MeV.
Lines are the results of the FAM emulator with 11 training points. Both are evaluated with 
${\rm Im}\,\omega=0.01$ MeV. The legend is identical to that of Fig.~\ref{fig:dy-gremulator}.
\label{fig:dy-lowlyingemulator}}
\end{figure}

Figure~\ref{fig:dy-lowlyingemulator} compares the strength distributions below 2.5 MeV on a logarithmic scale. 
The lowest-energy states are relatively well separated in energy from the other peaks, and well described by the emulator.
Table~\ref{table:Dycomparison} compares the energies and transition strengths for the lowest-energy $K^\pi=0^+$ states.
The FAM emulator reproduces at least four digits in the excitation energy, and thus is extremely good at locating the QRPA poles.
The proton quadrupole strength is related to the $B(E2)$ values between the ground-state rotational band and the $\beta$-vibrational band. The agreement between the proton quadrupole strength from the FAM and from the emulator is comparable to that for the isoscalar quadrupole strength, demonstrating that $B(E2)$ values can be obtained without explicitly using the proton operator in constructing the emulator.

\begin{table*}
\caption{
Excitation energies, isoscalar quadrupole strengths, and the proton quadrupole strengths for the lowest $K^\pi=0^+$ state of $^{156-180}$Dy isotopes.
The values obtained from the contour integration of the FAM are shown together with the differences between the FAM emulator and the FAM (RBM$-$FAM) for the low-lying states. \label{table:Dycomparison}}
\begin{ruledtabular}
\begin{tabular}{lllllll}
& \multicolumn{2}{c}{energy (MeV)} & \multicolumn{2}{c}{ISQ (fm$^4$)} &
  \multicolumn{2}{c}{proton ($e^2$ fm$^4$)} \\
& FAM & RBM$-$FAM & FAM  & RBM$-$FAM & FAM  & RBM$-$FAM  \\ \hline
$^{156}$Dy & 1.3116  &  \,\,\,\,\,0.0000002    & 929.51  &  \,\,\,\,\,0.00165 & 914.82 & \,\,\,\,\,0.00143  \\
$^{158}$Dy & 1.6613  &  \,\,\,\,\,0.0000293    & 426.42  &  \,\,\,\,\,0.01609 & 412.06 & \,\,\,\,\,0.03954  \\
$^{160}$Dy & 1.7018  &  \,\,\,\,\,0.0001240    & 2.0738  & $-0.00210$ & 2.7201 & $-0.00872$  \\
$^{162}$Dy & 1.6229  & $-0.0000241$  & 1.0276  & $-0.00279$ & 0.95370 & $-0.00055$  \\
$^{164}$Dy & 1.5476  &  \,\,\,\,\,0.0000584    & 0.71670 &  \,\,\,\,\,0.00041 & 0.74996 &  \,\,\,\,\,0.00167   \\
$^{166}$Dy & 1.4582  &  \,\,\,\,\,0.0000228    & 0.51051 & $-0.00163$ & 0.52355  & \,\,\,\,\,0.00152 \\
$^{168}$Dy & 1.4400  & $-0.0002733$  & 0.25568 & $-0.00064$ & 0.21260 &$-0.00330$  \\
$^{170}$Dy & 1.3998  &  \,\,\,\,\,0.0001134    & 0.31512 &  \,\,\,\,\,0.00003 & 0.47738  & $-0.00220$  \\
$^{172}$Dy & 1.3641  &  \,\,\,\,\,0.0000551    & 10.516  & $-0.00085$ & 11.060 & $-0.00523$   \\
$^{174}$Dy & 1.3093  &  \,\,\,\,\,0.0000025    & 114.54  & \,\,\,\,\,0.01341 & 101.53 & \,\,\,\,\,0.00981  \\
$^{176}$Dy & 1.2050  &  \,\,\,\,\,0.0000007    & 353.96  &  \,\,\,\,\,0.00130 & 296.02  & \,\,\,\,\,0.00081   \\
$^{178}$Dy & 1.0823  & $-0.0000022$  & 596.15  & \,\,\,\,\,0.00292 & 487.27 & \,\,\,\,\,0.00116   \\
$^{180}$Dy & 0.95161 & $-0.000000005$& 890.80  & $-0.00025$ & 714.59 & $-0.00030$
\end{tabular}
\end{ruledtabular}
\end{table*}

\section{Summary \label{sec:conclusion}}

We have used the RBM to introduce a reduced-order emulator for the FAM-QRPA. 
By taking the complex frequency of the applied external field as a training point, the emulator constructs an efficient subspace from a small number of FAM solutions.
This approach enables rapid evaluation of FAM amplitudes, strength functions, and QRPA eigenmodes without repeatedly solving the full linear-response equations.
Tests show that the RBM emulator accurately reproduces both giant-resonance strength distributions and low-energy collective excitations.
In contrast to iterative Arnoldi methods, the emulator allows one to selectively focus on a particular energy region, making it especially useful for low-energy dynamics.
The emulator also successfully describes imaginary QRPA solutions associated with shape instabilities, making it useful for calculations of collective vibrational inertia.
A realistic study of $\beta$ vibration in Dy isotopes shows that the emulator can reproduce QRPA energies and transition strengths with high fidelity, while reducing the computational cost of FAM-based strength-function and searches for low-energy poles by more than an order of magnitude.
Although there remains room for improvement in the way complex training energies are selected, we have established the RBM-based FAM emulator as an efficient and accurate tool for applications such as EDF parameter optimization, collective-inertia calculations, and global surveys of collective modes that require repeated QRPA calculations.

Preliminary results of the present work were reported in Refs.~\cite{Hinohara2023talk,Hinohara2025talk}. While this manuscript was being completed, Jin \textit{et al.}~\cite{PhysRevRes.7.043347} reported emulators for QRPA strength functions and derived observables, in which the elements of the reduced matrices are optimized against full-order QRPA calculations. The two approaches emulate along different directions: that work is trained on full-order calculations sampled over the EDF coupling constants, and represents the resulting strength function by a small number of effective poles; the present emulator is trained on FAM solutions sampled over the complex external-field energy at fixed EDF, and resolves the individual QRPA eigensolutions. The reduced matrices here are obtained by exact projection of the FAM equation, with no adjustable parameters, and reproduce the QRPA energies to $10^{-6}$--$10^{-3}$~MeV (Tables~\ref{table:24Mglowenergy} and \ref{table:spherical24Mg}) from a small number of training calculations. The same projection can be applied to the dependence on the EDF parameters, which we will address elsewhere.

\section*{Acknowledgments}
The authors are grateful to Dr. Pablo Giuliani, Kyle Godbey, Tong Li, Takashi Nakatsukasa, Witold Nazarewicz, Ante Ravli\'{c}, and Kouhei Washiyama for useful discussions.
The work of NH was supported by the JSPS KAKENHI (Grants No. JP19KK0343, No. JP20K03964, No. JP22H04569, No. JP25K07312, No. JP25K24695, No. JP26H01395, and No. JP26H02035) and by the JST ERATO Grant No. JPMJER2304.
NH acknowledges NUCLEI SciDAC-5 Collaboration under the U.S. Department of Energy.
XZ acknowledges support from the U.S. Department of Energy, Office of Science, Office of Nuclear Physics, under the FRIB Theory Alliance Award No. DE-SC0013617 and under the  STREAMLINE Collaboration Award No. DE-SC0024586 (Michigan State University) and the STREAMLINE 2 Collaboration Award No. DE-SC0026198. 
JE acknowledges support from the US Department of Energy, Office of Nuclear Physics under grant DE-FG02-97ER41019, and Office of Advanced Scientific Computing Research under grant DE-SC0023495. 
Numerical calculations were performed in part by using Miyabi-C supercomputer, provided by the Multidisciplinary Cooperative Research Program in the Center for Computational Sciences, University of Tsukuba.

\appendix

\section{\texorpdfstring{$PQ$ representation \label{sec:PQrepresentation}}{PQ representation}}

\subsection{QRPA}
If the stability matrix ${\cal S}$ in Eq.~(\ref{eq:stabilitymat}) is not fully positive definite, zero eigenvalues and/or pairs of imaginary eigenvalues appear.
One must use the $PQ$ representation to handle such solutions because the eigenvectors are not normalizable with Eq.~(\ref{eq:XYnormalization}). 
One then writes the QRPA equations in terms of the coordinate and momentum vectors $Q$ and $P$ as
\begin{align}
 {\cal S}{\cal V} = \Sigma_3 {\cal V}{\cal W} \,,
\end{align}
with
\begin{align}
 {\cal V} = \begin{pmatrix}  P & Q \\ -P^\ast & -Q^\ast \end{pmatrix}, \quad
 {\cal W} = \begin{pmatrix} 0 & -i M^{-1} \\ iM\Omega^2 & 0 \end{pmatrix} \,,
\end{align}
where $M_{ij}=M_i\delta_{ij}$ represents the inertial tensor and the vectors $Q$ and $P$ are connected with the $X$ and $Y$ vectors, if  $\Omega_\lambda$ is real and positive, through the relation
\begin{subequations}
\begin{align}
 Q^\lambda_{\mu\nu} &= \sqrt{\frac{1}{2M_\lambda\Omega_\lambda}} (X^\lambda_{\mu\nu} - Y^{\lambda\ast}_{\mu\nu}),\\
 P^\lambda_{\mu\nu} &= i\sqrt{\frac{ M_\lambda\Omega_\lambda}{2}} 
 ( X^\lambda_{\mu\nu} + Y^{\lambda\ast}_{\mu\nu}) \,.
\end{align}
\end{subequations}
The new vectors are normalized through the conditions 
\begin{align}
    {\cal V}^\dag \Sigma_3 {\cal V} = \Sigma_2, \quad
    {\cal V}\Sigma_2 {\cal V}^\dag = \Sigma_3 \,,
\end{align}
where
\begin{align}
 \Sigma_2 = \begin{pmatrix} 0 & -i \\ i & 0 \end{pmatrix} \,.
\end{align}

\subsection{FAM}

The FAM equation provides a formal relation between the FAM amplitudes and the QRPA eigenvectors in the $PQ$ representation \cite{Blaizot-Ripka,PhysRevC.92.034321}.
The matrix in Eq.~(\ref{eq:FAMeq}) can be decomposed: 
\begin{align}
[ {\cal S} - \omega \Sigma_3]^{-1} &=
{\cal V} [{\cal W} - \omega]^{-1} \Sigma_2 {\cal V}^\dag \,.
\end{align}
The FAM amplitudes are the following explicit linear combinations of QRPA eigenvectors:
\begin{subequations}
    \begin{align}
    X_{\mu\nu}(\omega) 
    &= \sum_{\lambda} \left[ 
\frac{(-i\omega P^{\lambda}_{\mu\nu} + M_{\lambda}\Omega_{\lambda}^2 Q^{\lambda}_{\mu\nu}) \langle Q_{\lambda}|\hat{F}|0\rangle}
 {\omega^2 - \Omega_{\lambda}^2} \right.
\nonumber \\
    &\quad + \left.
\frac{(M_{\lambda}^{-1}P_{\mu\nu}^{\lambda} + i\omega Q_{\mu\nu}^{\lambda}) \langle P_{\lambda}| \hat{F}|0\rangle}
 {\omega^2 - \Omega_{\lambda}^2} \right] \,,
    \\
Y_{\mu\nu}(\omega) &= 
     \sum_{\lambda} \left[
     \frac{(i\omega P^{\lambda\ast}_{\mu\nu} - M_{\lambda}\Omega_{\lambda}^2 Q^{\lambda\ast}_{\mu\nu}) \langle Q_{\lambda}|\hat{F}|0\rangle}
     {\omega^2 - \Omega_{\lambda}^2} \right.
 \nonumber \\
     &\quad + \left.
 \frac{( -M_{\lambda}^{-1}P^{\lambda\ast}_{\mu\nu} - i\omega Q^{\lambda\ast}_{\mu\nu}) \langle P_{\lambda}|\hat{F}|0\rangle}
 {\omega^2 - \Omega_{\lambda}^2} \right] \,,
\end{align} \label{eq:PQrep_FAMXY}
\end{subequations}
where the summation over $\lambda$ is performed only once for each pair of eigenvalues, avoiding double-counting.
The transition matrix elements in this representation are
\begin{subequations}
\begin{align}
 \langle Q_{\lambda}|\hat{F}|0\rangle &= Q^{\lambda\ast}\cdot F^{20} - Q^{\lambda}\cdot F^{02}\,, \\
 \langle P_{\lambda}|\hat{F}|0\rangle &= P^{\lambda\ast}\cdot F^{20} - P^{\lambda}\cdot F^{02}\,.
\end{align}
\end{subequations}
The FAM strength function is
\begin{align}
 S(\omega) &= 
 F^{20\ast}\cdot X(\omega) + F^{02^\ast}\cdot Y(\omega) \nonumber \\
 &=  
 \sum_{\lambda} \left[ 
 \frac{M_{\lambda}^{-1}|\langle P_{\lambda}|\hat{F}|0\rangle|^2 + M_{\lambda}\Omega_{\lambda}^2 |\langle Q_{\lambda}|\hat{F}|0\rangle|^2}
 {\omega^2 - \Omega_{\lambda}^2}  \right.
 \nonumber \\ 
 & \quad\quad\quad\quad + \left. \frac{\omega [QP]_{\lambda}}{\omega^2 - \Omega_{\lambda}^2} \right] \,,
 \label{eq:PQrep_strength}
\end{align}
where 
\begin{align}
[QP]_\lambda \equiv i\left( \langle Q_\lambda|\hat{F}|0\rangle^\ast \langle P_\lambda|\hat{F}|0\rangle
 - \langle P_\lambda|\hat{F}|0\rangle^\ast \langle Q_\lambda| \hat{F}|0\rangle \right) \,.
\end{align}

\subsection{RBM}

The RBM emulator is constructed in the same manner as discussed in Sec.~\ref{sec:RBM}; the 
FAM amplitudes $X(\omega_j)$ and $Y(\omega_j)$ are well defined for the system with zero- and imaginary-energy QRPA solutions.
The collective Hamiltonian (\ref{eq:collH}) is also constructed in the same way.  Zero and imaginary eigenvalues may be present, however, especially when the original stability matrix ${\cal S}$ is not fully positive definite.
Even when it is positive definite, numerical error may cause imaginary eigenvalues to appear in the emulator. 
If zero and imaginary eigenvalues exist, the eigenvectors of the QRPA equation in the reduced-order space 
(\ref{eq:G}) are not normalizable through Eqs.~(\ref{eq:Ginv}).
In such a situation, Eq.~(\ref{eq:ung}) should be written as
\begin{subequations}
\begin{align}
 {\sf U}\bar{\sf N}^{-\frac{1}{2}} {\sf G} &= 
 \begin{pmatrix}
 \tilde{\sf U}^{(1)} & -i \tilde{\sf V}^{(2)\ast} \\ \tilde{\sf V}^{(1)} & -i \tilde{\sf U}^{(2)\ast}
 \end{pmatrix}, \\
 {\sf G}^{-1}\bar{\sf N}^{-\frac{1}{2}} {\sf U}^\dag &= 
 \begin{pmatrix}
 \tilde{\sf U}^{(3)\dag} & \tilde{\sf V}^{(3)\dag} \\ -i\tilde{\sf V}^{(4)T} & -i \tilde{\sf U}^{(4)T}
 \end{pmatrix} \,.
 \end{align}
\end{subequations}
The response function and the coefficients of the FAM amplitudes in the RBM are given by
\begin{widetext}
\begin{align}
 [{\sf H}-\omega{\sf N}]^{-1} =
 \begin{pmatrix}
    \tilde{\sf U}^{(1)}(\tilde{\sf \Omega}-\omega)^{-1}\tilde{\sf U}^{(3)\dag} + \tilde{\sf V}^{(2)\ast}(\tilde{\sf \Omega} + \omega)^{-1}\tilde{\sf V}^{(4)T} & 
    \tilde{\sf U}^{(1)}(\tilde{\sf \Omega}-\omega)^{-1}\tilde{\sf V}^{(3)\dag} + \tilde{\sf V}^{(2)\ast}(\tilde{\sf \Omega} + \omega)^{-1}\tilde{\sf U}^{(4)T} \\
    \tilde{\sf V}^{(1)}(\tilde{\sf \Omega}-\omega)^{-1}\tilde{\sf U}^{(3)\dag} + \tilde{\sf U}^{(2)\ast}(\tilde{\sf \Omega} + \omega)^{-1}\tilde{\sf V}^{(4)T} & 
    \tilde{\sf V}^{(1)}(\tilde{\sf \Omega}-\omega)^{-1}\tilde{\sf V}^{(3)\dag} + \tilde{\sf U}^{(2)\ast}(\tilde{\sf \Omega} + \omega)^{-1}\tilde{\sf U}^{(4)T}
 \end{pmatrix}\,, 
\label{eq:PQrep_H-wN}
\end{align}
\begin{align}
\begin{pmatrix} a_i(\omega) \\ b_i(\omega) \end{pmatrix} =
- \begin{pmatrix} 
\tilde{\sf U}^{(1)}     (\tilde{\sf \Omega} - \omega)^{-1} \tilde{\sf S}^{(3)\ast} +
\tilde{\sf V}^{(2)\ast} (\tilde{\sf \Omega} + \omega)^{-1} \tilde{\sf S}^{(4)\ast} \\
\tilde{\sf V}^{(1)}     (\tilde{\sf \Omega} - \omega)^{-1} \tilde{\sf S}^{(3)\ast} + 
\tilde{\sf U}^{(2)\ast} (\tilde{\sf \Omega} + \omega)^{-1} \tilde{\sf S}^{(4)\ast}
\end{pmatrix} \,, 
\label{eq:PQrep_ab}
\end{align}
\end{widetext}
with
\begin{subequations}
\begin{align}
\tilde{\sf S}^{(3)} &= \tilde{\sf U}^{(3)T} {\sf S} + \tilde{\sf V}^{(3)T} {\sf S}^{\prime\ast}, \\ 
\tilde{\sf S}^{(4)} &= \tilde{\sf V}^{(4)\dag} {\sf S} + \tilde{\sf U}^{(4)\dag} {\sf S}^{\prime\ast} \,.
\end{align}
\end{subequations}
By substituting Eq.~(\ref{eq:PQrep_H-wN}) into Eq.~(\ref{eq:emulatorstrength}), the emulated strength function becomes 
\begin{align}
S(\hat{F},\omega) &= - \tilde{\sf S}^{(1)} (\tilde{\sf \Omega} - \omega)^{-1} \tilde{\sf S}^{(3)\ast}
- \tilde{\sf S}^{(2)} ( \tilde{\sf \Omega} + \omega)^{-1} \tilde{\sf S}^{(4)\ast} \,,
\label{eq:PQrep_strengthemulator}
\end{align}
where 
\begin{subequations}
\begin{align}
    \tilde{\sf S}^{(1)} &= \tilde{\sf U}^{(1)T} {\sf S} + \tilde{\sf V}^{(1)T} {\sf S}^{\prime\ast}, \\
    \tilde{\sf S}^{(2)} &= \tilde{\sf V}^{(2)\dag} {\sf S} + \tilde{\sf U}^{(2)\dag} {\sf S}^{\prime\ast} \,.
\end{align}
\end{subequations}
By comparing Eqs.~(\ref{eq:PQrep_strength}) and (\ref{eq:PQrep_strengthemulator}), we obtain 
\begin{subequations}
\begin{align}
\frac{|\langle\tilde{P}_{\lambda}|\hat{F}|0\rangle|^2}{M_{\lambda}\tilde{\Omega}_{\lambda}} + M_{\lambda}\tilde{\Omega}_{\lambda}|\langle\tilde{Q}_{\lambda}|\hat{F}|0\rangle|^2 &\equiv \tilde{\sf S}_{\lambda}^{(1)}\tilde{\sf S}_{\lambda}^{(3)\ast} +
\tilde{\sf S}_{\lambda}^{(2)}\tilde{\sf S}_{\lambda}^{(4)\ast}, \\
[\tilde{Q}\tilde{P}]_{\lambda} &\equiv \tilde{\sf S}_{\lambda}^{(1)}\tilde{\sf S}_{\lambda}^{(3)\ast} -
\tilde{\sf S}_{\lambda}^{(2)}\tilde{\sf S}_{\lambda}^{(4)\ast} \,.
\end{align} \label{eq:PQrep_strengthPQ}
\end{subequations}
The FAM emulator amplitudes, in explicit form, are 
\begin{subequations}
\begin{align}
 {\sf X}(\omega) &= -\tilde{\sf X}^{(1)}(\tilde{\sf \Omega} - \omega)^{-1} \tilde{\sf S}^{(3)\ast}
 - \tilde{\sf Y}^{(2)\ast} (\tilde{\sf \Omega} + \omega)^{-1}\tilde{\sf S}^{(4)\ast} \,, \\
 {\sf Y}(\omega) &= -\tilde{\sf Y}^{(1)}(\tilde{\sf \Omega} - \omega)^{-1} \tilde{\sf S}^{(3)\ast}
 - \tilde{\sf X}^{(2)\ast} (\tilde{\sf \Omega} + \omega)^{-1}\tilde{\sf S}^{(4)\ast} \,,
\end{align}
\label{eq:PQrep_emulatorXY}
\end{subequations}
where 
\begin{align}
\left.
\begin{aligned}
\tilde{\sf X}^{(k)\lambda}_{\mu\nu} &= \sum_{i=1}^{2n}
   \left[ X_{\mu\nu}(\omega_i)\tilde{\sf U}^{(k)}_{i\lambda}
        + Y^{*}_{\mu\nu}(\omega_i)\tilde{\sf V}^{(k)}_{i\lambda} \right], \\
\tilde{\sf Y}^{(k)\lambda}_{\mu\nu} &= \sum_{i=1}^{2n}
   \left[ Y_{\mu\nu}(\omega_i)\tilde{\sf U}^{(k)}_{i\lambda}
        + X^{*}_{\mu\nu}(\omega_i)\tilde{\sf V}^{(k)}_{i\lambda} \right],
\end{aligned}
\right\}
\quad (k = 1, 2) \label{eq:PQrep_Emulator_QRPA_XY}
\end{align}
When the operator $\hat{F}$ is Hermitian, $\langle \tilde{P}_\lambda|\hat{F}|0\rangle$ and  $\langle \tilde{Q}_\lambda|\hat{F}|0\rangle$ are imaginary, and by comparing Eqs.~(\ref{eq:PQrep_FAMXY}) and (\ref{eq:PQrep_emulatorXY}), we get 
\begin{subequations}
\begin{align}
 \tilde{\sf X}_{\mu\nu}^{(2)\lambda}\tilde{\sf S}_\lambda^{(4)} = \tilde{\sf X}_{\mu\nu}^{(1)\lambda}\tilde{\sf S}_\lambda^{(3)\ast}, \\
 \tilde{\sf Y}_{\mu\nu}^{(2)\lambda}\tilde{\sf S}_\lambda^{(4)} = \tilde{\sf Y}_{\mu\nu}^{(1)\lambda}\tilde{\sf S}_\lambda^{(3)\ast} \,.
\end{align}
\end{subequations}
In many applications, one can choose $\tilde{\sf Q}^\lambda$ to be real and $\tilde{\sf P}^\lambda$ to be imaginary.
In this case $\langle \tilde{Q}_{\lambda}|\hat{F}|0\rangle=0$.
The matrix elements are given by
\begin{subequations}
\begin{align}
\tilde{\sf Q}^\lambda_{\mu\nu} &= 
-i \frac{ (\tilde{\sf X}_{\mu\nu}^{(1)\lambda} - \tilde{\sf Y}_{\mu\nu}^{(1)\lambda}) \tilde{\sf S}^{(3)\ast}_\lambda} 
 {\langle \tilde{P}_\lambda|\hat{F}|0 \rangle}, \label{eq:Emulator_QRPA_Q} \\
\tilde{\sf P}^\lambda_{\mu\nu} &=
 \frac{ (\tilde{\sf X}_{\mu\nu}^{(1)\lambda} + \tilde{\sf Y}_{\mu\nu}^{(1)\lambda}) \tilde{\sf S}^{(3)\ast}_\lambda} 
 {\langle \tilde{P}_\lambda|\hat{F}|0 \rangle}. \label{eq:Emulator_QRPA_P} 
\end{align}
\end{subequations}
From Eq.~(\ref{eq:PQrep_strengthPQ}), we have $\tilde{\sf S}^{(1)}_\lambda \tilde{\sf S}^{(3)\ast}_\lambda
= \tilde{\sf S}^{(2)}_\lambda \tilde{\sf S}^{(4)\ast}_\lambda$ ($[\tilde{Q}\tilde{P}]_\lambda=0$),
and the transition strength in the $PQ$ representation is given by 
\begin{align}
|\langle\tilde{P}_{\lambda}|\hat{F}|0\rangle|^2 &=
2M_\lambda \tilde{\Omega}_\lambda 
\tilde{\sf S}_{\lambda}^{(1)}\tilde{\sf S}_{\lambda}^{(3)\ast} \,.
\end{align}

\section{Iterative Arnoldi method for QRPA \label{sec:Arnoldi}}

In the iterative Arnoldi method, we construct a series of Arnoldi vectors 
$|Z_i\rangle=(\bm{X}_i, \bm{Y}_i)^T$ 
and $|Z_i'\rangle = (\bm{Y}_i^\ast, \bm{X}_i^\ast)^T$
in the following way.
A pair of first Arnoldi vectors is constructed from an external-field operator $\hat{F}$
\begin{align}
|Z_1\rangle = 
\begin{pmatrix}
 \bm{X}_1 \\ \bm{Y}_1
\end{pmatrix}
= \frac{1}{ \sqrt{ \bm{F}^{20\dag}\cdot \bm{F}^{20}}}
\begin{pmatrix} 
\bm{F}^{20} \\ \bm{0}
\end{pmatrix}, 
\quad
|Z'_1 \rangle = 
\begin{pmatrix} \bm{Y}^\ast_1 \\ \bm{X}^\ast_1
\end{pmatrix} \,.
\end{align}
Then we compute the product of the 
QRPA matrix and the Arnoldi vector
\begin{align}
|W_n\rangle &= 
\begin{pmatrix} \bm{W}_n \\ \bm{W}'_n \end{pmatrix}
=  
\begin{pmatrix} A & B \\ -B^\ast & -A^\ast 
\end{pmatrix}
\begin{pmatrix} \bm{X}_n \\ \bm{Y}_n
\end{pmatrix}
\nonumber \\
&= 
\begin{pmatrix}
 \bm{\delta H}^{20}(\bm{X}_n,\bm{Y}_n)
+ \overline{E} \bm{X}_n \\
 -\bm{\delta H}^{02}(\bm{X}_n,\bm{Y}_n) - \overline{E} \bm{Y}_n 
\end{pmatrix}\,,
\end{align}
where we can avoid the multiplication of $A$ and $B$ matrices as described in Ref.~\cite{PhysRevC.81.034312}.
The iterative Arnoldi method requires the 
orthogonalization of the vector with all the previous Arnoldi vectors.
The unnormalized $(n+1)$-th Arnoldi vector is 
given by $|W_n\rangle$ after
orthogonalization with all 
the previously obtained $|Z_i\rangle$ and 
$|Z'_i\rangle$ vectors as
\begin{align}
|\tilde{Z}_{n+1}\rangle
=\begin{pmatrix} \tilde{\bm{X}}_{n+1} \\ \tilde{\bm{Y}}_{n+1}
\end{pmatrix}
= |W_{n}\rangle
 - \sum_{i=1}^{n} a_{in} |Z_i\rangle
 + \sum_{i=1}^n b_{in} |Z'_i\rangle\,, \label{eq:Zn+1}
\end{align}
where the coefficients are given by
\begin{align}
a_{in} &= \langle Z_i|\Sigma_3|W_n\rangle
= 
\bm{X}^\dag_i \cdot \bm{W}_n 
-
\bm{Y}^\dag_i \cdot \bm{W}'_n, \\
b_{in} &= \langle Z'_i|\Sigma_3 |W_n\rangle
=
\bm{Y}^T_i \cdot \bm{W}_n - \bm{X}^T_i\cdot\bm{W}'_n \,.
\end{align}
The normalization of the vector $|\tilde{Z}_{n+1}\rangle$ is given by
\begin{align}
\tilde{N}_{n+1} = 
\langle \tilde{Z}_{n+1}|\Sigma_3 |\tilde{Z}_{n+1}\rangle
= 
\tilde{\bm{X}}_{n+1}^\dag\cdot \tilde{\bm{X}}_{n+1}
-
\tilde{\bm{Y}}_{n+1}^\dag\cdot\tilde{\bm{Y}}_{n+1} \,.
\end{align}
The normalized $(n+1)$-th Arnoldi vectors are given by
\begin{align}
|Z_{n+1}\rangle 
&= \frac{1}{\sqrt{ \tilde{N}_{n+1}}}
\begin{pmatrix}
\tilde{\bm{X}}_{n+1} \\
\tilde{\bm{Y}}_{n+1}
\end{pmatrix}, \\
|Z_{n+1}'\rangle 
&= \frac{1}{\sqrt{ \tilde{N}_{n+1}}}
\begin{pmatrix}
\tilde{\bm{Y}}_{n+1}^\ast \\
\tilde{\bm{X}}_{n+1}^\ast
\end{pmatrix}\,,
 \end{align}
 for $\tilde{N}_{n+1}>0$, and
 \begin{align}
|Z_{n+1}\rangle 
&= \frac{1}{\sqrt{ -\tilde{N}_{n+1}}}
\begin{pmatrix}
\tilde{\bm{Y}}^\ast_{n+1} \\
\tilde{\bm{X}}^\ast_{n+1}
\end{pmatrix}, \\
 |Z'_{n+1}\rangle 
&= \frac{1}{\sqrt{ -\tilde{N}_{n+1}}}
\begin{pmatrix}
\tilde{\bm{X}}_{n+1} \\
\tilde{\bm{Y}}_{n+1}
\end{pmatrix}\,,
\end{align}
for $\tilde{N}_{n+1}<0$.

The matrices that orthogonalize the new Arnoldi vectors with the previous ones in Eq.~(\ref{eq:Zn+1}) correspond to the Krylov-space QRPA
matrix elements, and the QRPA equation in the Krylov space is given by
\begin{align}
\begin{pmatrix} a & b \\ -b^\ast & -a^\ast
\end{pmatrix}
\begin{pmatrix} x^k \\ y^k \end{pmatrix}
= 
\Omega_k
\begin{pmatrix} x^k \\ y^k \end{pmatrix} \,.
\end{align}
The eigenvectors in the Krylov space are 
normalized in the same way for real eigenvalues 
\begin{align}
{\mathscr X}^\dag \Sigma_3 {\mathscr X}&= \Sigma_3, \quad
{\mathscr X} \Sigma_3 {\mathscr X}^\dag = \Sigma_3, \\
 {\mathscr X} &= \begin{pmatrix} x & y^\ast \\ 
 y & x^\ast \end{pmatrix} \,.
\end{align}
The QRPA eigenvectors ($\bm{X}^k, \bm{Y}^k), (k=1,\cdots, d$), where $d$ is the number of the Arnoldi basis computed, are given by
\begin{align}
 \bm{X}^k &=
 \sum_{l=1}^d \bm{X}_l x_l^k
 + \bm{Y}_l^\ast y_l^{k\ast}, \\
 \bm{Y}^k &=
 \sum_{l=1}^d \bm{Y}_l x_l^{k\ast}
 + \bm{X}_l^\ast y_l^{k}, 
\end{align}
where $l$ runs the indices for the Arnoldi vectors, and the $\bm{X}_l$ and $\bm{Y}_l$ in the right hand side are the Arnoldi vectors.

It is also straightforward to calculate the QRPA eigenvectors in PQ representation, $\bm{Q}^k, \bm{P}^k$.

\bibliography{rbm_fam}

\end{document}